\documentclass[varenna]{cimento}

\usepackage{graphicx}  % got figures? uncomment this
\title{Density Functional Theory (DFT) for atomic nuclei: a simple
introduction}

\author{G. Col\`o}

\institute{Dipartimento di Fisica, Universit\`a
degli Studi di Milano,\\ via Celoria 16, 20133 Milano, Italy\\
and INFN, Sezione di Milano, via Celoria 16, 20133 Milano, Italy}

\begin{document}

\maketitle

\begin{abstract}
The present contribution does not aim at replacing the huge and often excellent literature
on DFT for atomic nuclei, but tries to 
provide an updated introduction to this topic. The goal would be, ideally, to help a
fresh M.Sc. or Ph.D. student (or a researcher from other fields) to become acquainted 
with some basic concepts, and then move to the specialized textbooks or papers with some
ability for orienteering. We first introduce the basics of DFT, and show the difference 
with the ``na\"ive'' mean-field theory, that is doomed to fail as a model
even in the simple case of uniform nuclear matter. We introduce the Energy Density Functionals 
(EDFs) that are used in nuclear structure, with few examples of their applications. 
The concepts of symmetry 
breaking and restoration are briefly discussed. We also include an introduction to the time-dependent extension
of DFT that, so far, has been implemented essentially only in the adiabatic approximation and has been applied mainly 
to the study 
of nuclear vibrations. 
With this material,
we hope that any reader is able to deal with the texts that go deeper into each of
the topics, having understood that DFT is probably the best compromise in nuclear
structure theory between simplicity, accuracy, and broad range of applicability.
\end{abstract}

\section{Introduction}

\begin{figure}[b!]
\centering\includegraphics[width=0.9\textwidth]{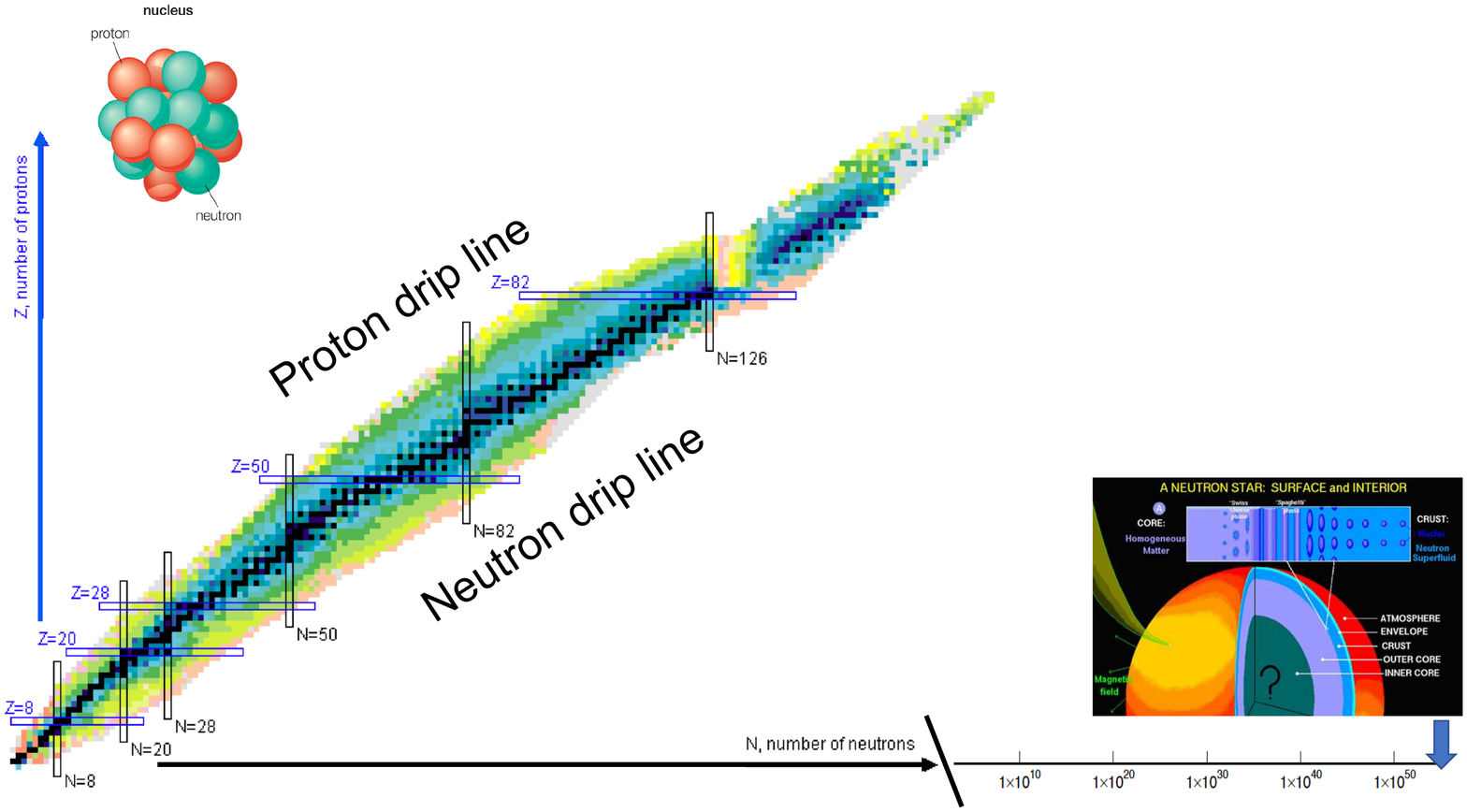}
\caption{Overall view of the nuclear chart. In the right part, 
a schematic picture of a neutron star (taken from D. Page), is displayed.}
\label{fig1}
\end{figure}

Atomic nuclei are strongly correlated, self-bound quantum systems that are still
capturing the interest of so many scientists, more than a century after their 
discovery, for a variety of reasons. 

The nuclear chart, that is a two-dimensional
arrangement of nuclei on a plane where the number of neutrons, $N$, and the number of
protons, $Z$, are the $x-$ and $y-$axis, respectively, has still broad territories to
be explored. A schematic view from the web (https://www.nndc.bnl.gov/chart/) is displayed 
in Fig. \ref{fig1}. Every year, a large number of new nuclei, that amount to $\approx$ 20-30 in the 
last decade, are being discovered (cf. \cite{Thoennessen:2017} and references therein).
The limit of existence for neutron-rich or neutron-deficient nuclei (so-called drip lines,
beyond which nuclei are unbound with respect to neutron and proton emission, respectively), 
and the search for superheavy elements, are the highlights of this exploration. 
 
At the same time, the nuclear physics and nuclear astrophysics communities are striving
to grasp some understanding of compact objects like neutron stars, that are extreme
forms of nuclear matter. While ordinary nuclei display densities around the so-called
saturation density $\rho_0$ = 0.16 fm$^{-3}$ (see below) and are at, or lie close to, zero
temperature, different conditions may be realised in stars. The inner core of neutron
stars contains matter in unknown conditions that may also correspond to a quark-deconfined
phase, while the crust is made up with nucleons whose density encompasses a broad range
from around 1/3 to 10$^{-3}$ times the saturation density \cite{CH.08}. 
Neutron stars are believed to contain 
$\approx$ 10$^{55-56}$ neutrons, and as such are displayed in the lower-right corner of
Fig. \ref{fig1}.

The new input from nuclei with large neutron-proton asymmetry, or superheavy isotopes, or 
exotic forms of neutron matter in stars, has changed and is still changing
our basic understanding of nuclear structure. 
Consequently, there is certainly a strong need of up-to-date introductions to several topics. 

The present lecture has mainly a theoretical content. Nuclear theory is moving
ahead quite rapidly in recent years. Many groups are pursuing the idea that the goal
should be deriving nuclear properties from (the low-energy limit of) Quantum Chromo
Dynamics (QCD). Genuine lattice calculations with explicit quark degrees of freedom
have still serious troubles to reproduce basic properties like 
the binding energy of the few-nucleon systems (two- and three-body systems turn out to be 
unbound while $^4$He is severely underbound \cite{Inoue:2013}). A different strategy 
consists in using an {\em effective} realization of the QCD Lagrangian, based on chiral 
symmetry as originally proposed by S. Weinberg \cite{Weinberg:1979}. Chiral Effective Field 
Theory (EFT) gives rise to a family of model Lagrangians, all based on correct symmetries
and separation of energy scales, that can be used in connection with many-body methods like 
lattice simulations, coupled cluster, renormalization group approaches, or Green's function
methods. These so-called {\em ab initio} methods are continuously extending their range of 
applicability; nonetheless applying them to heavy nuclei, or highly excited states, is still
too demanding \cite{RevModPhys.81.1773}.  More importantly, the quality of the results still depends on the specific
Lagrangian \cite{Gandolfi2016,Simonis2017}. 

As a summary, deriving nuclear properties directly from QCD is still a long-term project.
{\em Ab initio} is a wording that may be used in connection with many-body
methods that are in principle exact, 
not only in connection with chiral Lagrangians but also 
when a phenomenological nucleon-nucleon (NN) interaction is 
employed. In general, the applicability of these approaches has serious limitations in mass
number and excitation energy. 

In keeping also with the rich variety of the nuclear 
phenomena, one can easily understand the absence of a nuclear ``standard model''. 
In the current volume, the reader can browse through different
lectures and judge directly about the pros and cons of models like the nuclear Shell
Model, the Cluster Model, and the algebraic approaches; all these models necessarily 
include a fair amount of phenomenological input. DFT lies somehow in between the
purely phenomenological models and those that aim at starting from QCD. It can be said
in its own way to be an {\em ab initio} theory, because it is rooted in the Hohenberg-Kohn theorem
(cf. below). There is not, so far, a systematic connection with an underlying theory
but many attempts are promising. As we shall argue in this lecture, 
a reasonable account of experimental data can be achieved in a quite transparent and
economic way. Analogies with electronic systems can be traced, having in mind the
development of unified methods for many-fermion systems.  

The outline of the present contribution is the following. We discuss the basics of DFT, mainly
with reference to the electronic systems, in Sec. \ref{sec:basics}. In the nuclear case, 
we remind first the reader about the evidences for independent particle motion that may
lead to the assumption of the validity of simple HF (Sec. \ref{sec:HF}), and we then
introduce uniform nuclear matter (Sec. \ref{sec:matter}) and use it as a playground
to show the failure of na\"ive HF (Sec. \ref{sec:MF}). We advocate the need for density-dependent 
interactions that are merely generators of energy functionals, and we discuss the functionals
which are currently used in Sec. \ref{sec:examples}. Few examples of DFT calculations for
the ground-state properties are mentioned in Sec. \ref{sec:examples_gs}. We then move to
the concepts of intrinsic density and symmetry breaking in Sec. \ref{sec:intrinsic} and Sec. \ref{sec:sym}, 
respectively. Sec. \ref{sec:TD} is devoted to the extension of DFT to the time-dependent case, with 
one illustrative example in Sec. \ref{sec:examples_RPA}. Finally, we mention the limitations
of DFT in Sec. \ref{sec:limitations}, and we draw some conclusions in Sec. \ref{sec:conclu}.

Many suggestions for further reading are given along the text. Nevertheless, we outline
some of them here, for the reader's convenience. There are standard textbooks that
constitute a useful starting point for the study of nuclear structure along the line of this 
paper \cite{RS.80}. If the reader needs a recent, more introductory text, Ref. 
\cite{Bertulani} is an option. The literature on DFT in condensed matter is huge,
and yet we can single out Refs. \cite{ParrYang,DG.90,Gross:1996}. Recent, and relatively
short, papers on the perspectives for electronic DFT can be found \cite{Burke.12,Becke.14}.
DFT in nuclear physics
has been first reviewed in \cite{BHF.03}, but the reader can also profit from 
Refs. \cite{LNP.641,LNP.879}. There are many lectures available on the web, and
an ambitious and interesting attempt to introduce the subject of DFT in nuclear
physics can be found on the archive \cite{Doba:archive}. 

Last but not least, we limit ourselves here to a nonrelativistic treatment. Excellent
papers that introduce and explain the relativistic (or covariant) nuclear DFT
are available \cite{VALR.05,NVR.11}.

\section{Basics on DFT for electronic systems}\label{sec:basics}

We assume we are concerned with a quantum many-fermion system governed by the
Hamiltonian
\begin{equation}\label{eq:H1}
H = \sum_{i=1}^N -\frac{\hbar^2}{2m}\nabla_i^2 + \frac{1}{2}\sum_{i \ne j = 1}^N
V(i,j) + \sum_{i=1}^N v_{\rm ext}(i),
\end{equation}
where the first term is the kinetic energy of the $N$ fermions having mass $m$,
the second term is a two-body interaction in which $i$ is a shorthand notation for
the space coordinate, $\vec r_i$, spin coordinate and any further degree of freedom
(e.g. isospin), and the last term is a possible external potential. In the case of
$N$ electrons interacting with the external field of $M$ ions (labelled by $\alpha$, and
having charge $Z_\alpha$ and associated coordinate $\vec R_\alpha$), this Hamiltonian becomes
\begin{equation}
H = \sum_{i=1}^N -\frac{\hbar^2}{2m}\nabla_i^2 + \frac{1}{2} \sum_{i \ne j = 1}^N 
\frac{e^2}{\vert \vec r_i - \vec r_j \vert} + \sum_{i=1}^N \sum_{\alpha=1}^M 
\frac{Z_\alpha e^2}{\vert \vec r_i - \vec R_\alpha \vert},
\end{equation}
where the interactions between the ions are not written for the sake
of simplicity. 
The many-body problem associated with this Hamiltonian cannot be solved
exactly, even if the Coulomb force is well known, and even if one can invoke the
Born-Oppenheimer approximation and reduce the whole problem to the electronic problem,
for fixed ion positions.

Many strategies have been proposed to solve the many-electron problem. Among them,
to the opinion of many, DFT stands out for its conceptual elegance and simplicity (that, most
likely, should be a feature of physics as a whole). Its foundation 
lies in the theorem that bears the name of Hohenberg and Kohn (HK)
and that was introduced in their seminal paper \cite{HK.64}. In short, the theorem
states that the total energy of the system described by (\ref{eq:H1}), for any 
external potential $v_{\rm ext}$, can be written as a {\em functional} of the fermion density
$\rho (\vec r)$:
\begin{equation}\label{eq:HKtheorem}
E_{v_{\rm ext}}[\rho] = \langle \Psi \vert \hat T + \hat V + \hat v_{\rm ext} \vert \Psi \rangle =
F[\rho] + \int d^3r\ v_{\rm ext}(\vec r)\rho(\vec r).
\end{equation}
The first equality is just the definition of total energy, while the second equality
defines the functional: the contribution of the external potential is singled out and
the {\em universal functional} $F$ is introduced. The functional $E$, Eq. 
(\ref{eq:HKtheorem}), has a minimum at 
the exact ground-state density where it becomes equal to the exact energy. It is
hard to underestimate the value of Eq. (\ref{eq:HKtheorem}), because it tells us that
the whole information about the ground-state of the system is contained in the 
one-body density $\rho$ that is a real function of three coordinates, and we do not
need the whole wave function $\Psi$, that instead is a complex function of 
3$N$ coordinates! 

The proof of the theorem can be found in the original paper \cite{HK.64}. We advise the reader 
to go through it; it is quite instructive, although limited to the specific case in which 
the ground-state is not degenerate. Extensions of the theorem to the case of degenerate ground states, 
spin-polarized systems, finite temperature etc.\footnote{The list is not exhaustive.}, can be found in 
the literature (see, e.g., \cite{DG.90}). 

The real drawback of the 
HK theorem is that its proof is merely a proof of existence of the universal
functional $F$. It is not a constructive proof, so that the appropriate strategy to build this
functional remains an open problem. A step forward in this direction is represented by
the Kohn-Sham (KS) scheme \cite{KS.65}, in which 
it is assumed that the density $\rho$ can be represented in terms
of so-called auxiliary single-particle wave functions (orbitals) $\phi_j(\vec r)$, that is,
\begin{equation}
\rho(\vec r) = \sum_j \vert \phi_j(\vec r) \vert^2.
\end{equation}
The index $j$ labels the orbitals. The meaning
of the word ``auxiliary'' should be clarified here. The orbitals 
may be looked at as
a formal artifact. This does not exclude that, in specific cases, the orbitals turn out to be 
a fairly good approximation of the actual single-particle wave functions. 
Within such framework, 
the total kinetic energy is written in the usual way,
\begin{equation}\label{eq:Ekin}
T = \langle \Psi \vert \hat T \vert \Psi \rangle 
= \sum_j \langle j \vert \frac{-\hbar^2}{2m}\nabla^2 \vert j \rangle 
= \sum_j \int d^3r\ \phi_j^*(\vec r) \left(
- \frac{\hbar^2}{2m}\nabla^2 \right) \phi_j(\vec r),
\end{equation}
and 
the direct Coulomb energy (classical, or
Hartree, energy) can be also singled out:
\begin{equation}\label{eq:EHartree}
E_{\rm Hartree} = \frac{e^2}{2}\int d^3r\ d^3r^\prime\ \frac{\rho(\vec r)\rho(\vec r^\prime)}
{\vert \vec r - \vec r^\prime \vert}.
\end{equation}
In this way, the functional (\ref{eq:HKtheorem}) is re-written as
\begin{equation}\label{eq:KSfunctional}
E_{\rm KS} = T + E_{\rm Hartree} + E_{\rm xc} + \int d^3r\ v_{\rm ext}(\vec r)\rho(\vec r),
\end{equation}
where now the part to be built is the so-called ``exchange-correlation'' part or
$E_{\rm xc}$. The minimization of this latter functional with respect to the density should
be carried out with the constraint that the orbitals are orthonormal. Therefore, the 
equations for the orbitals are obtained from
\begin{equation}\label{eq:minKS}
\frac{\delta}{\delta \rho(\vec r)}\left( E_{\rm KS} - \varepsilon 
\int d^3r^\prime\ \phi_j^*(\vec r^\prime) 
\phi_j(\vec r^\prime) \right) = 0,
\end{equation}
where the usual symbol $\delta$ is introduced for the functional derivative. Elementary 
introductions to functional derivatives can be found in several textbooks (see, e.g., Sec. 3.2 of 
\cite{BCOR.04}). Eq. (\ref{eq:minKS}), together with Eqs. (\ref{eq:KSfunctional}), 
(\ref{eq:Ekin}) and (\ref{eq:EHartree}), leads to the famous Kohn-Sham equations
\begin{equation}\label{eq:KS}
\left( -\frac{\hbar^2}{2m}\nabla^2 + \frac{e^2}{2}\int d^3r^\prime\ 
\frac{\rho(\vec r^\prime)}{\vert \vec r - \vec r^\prime \vert} 
+ \frac{\delta E_{\rm xc}}{\delta\rho(\vec r)}
+ v_{\rm ext}(\vec r) 
\right) \phi_j(\vec r) 
= \left( -\frac{\hbar^2}{2m}\nabla^2 + v_{KS} \right) \phi_j(\vec r)
= \varepsilon \phi_j(\vec r),
\end{equation}
where we have labelled the total effective Kohn-Sham potential by $v_{KS}$, and
where the quantities $\varepsilon$ are seen to be the energies associated with the
auxiliary orbitals. The reader should note that the total energy is {\em not} the sum
of these auxiliary energies (this is left as an exercise). We also wish to stress
that the variation with respect to $\rho$ in Eq. (\ref{eq:minKS}) has in fact been turned 
into a variation with respect to the orbitals $\phi_j^*$. 

\begin{figure}[h!]
\centering\includegraphics[width=0.7\textwidth]{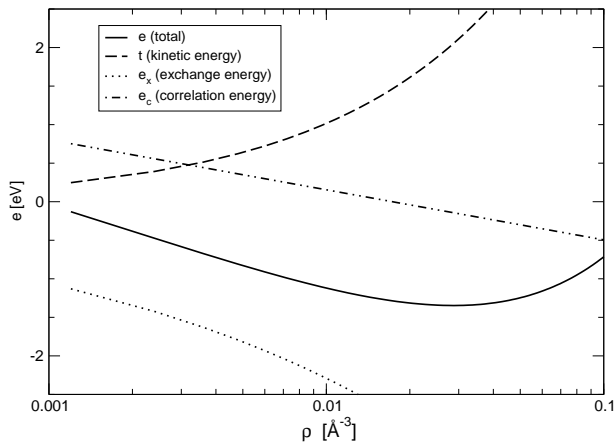}
\caption{Energy per particle of the uniform electron gas as a function of the density, within
the LDA as discussed in the main text. The curves correspond to the result 
of Eq. (\ref{eq:LDA_high_dens}) and to the three contributions that appear therein.}
\label{fig2}
\end{figure}

As for the choice for $E_{\rm xc}$, the 
simplest possible option is to calculate it 
in a {\em uniform} system where 
translational invariance makes the exact calculation of the total energy feasible. In fact,
such a calculation 
in the limit of high electron densities is analytic and can be found in textbooks (see, e.g., Secs. 3 and 
12 of \cite{Fetter} or Sec. 12.3 of \cite{Mattuck}). The total energy per
particle $\frac{E}{N}$ reads, 
\begin{equation}\label{eq:LDA_high_dens}
\frac{E}{N} = \frac{e^2}{2a_0}\left( \frac{2.21}{r_s^2} - \frac{0.916}{r_s} +
0.0622\ {\rm ln}r_s - 0.094 + \ldots \right) = t + e_{\rm x} + e_{\rm c}.
\end{equation}
One usually defines 
$a_0=\frac{\hbar^2}{me^2}$ as the Bohr radius, $r_0$ by $\frac{1}{\rho}=\frac{4}{3}\pi r_0^3$ so that it is
essentially the interparticle spacing, and $r_s=\frac{r_0}{a_0}$. In the latter equation, 
the dots represent terms of the order $O(r_s{\rm ln}r_s)$, and other subleading terms that
become negligible at high density, that is, small $r_s$.
The first term in brackets in Eq. (\ref{eq:LDA_high_dens}) 
is the kinetic energy per particle $t$, that goes like $\rho^{2/3}$. In this specific case, the 
Hartree and external potential contributions cancel exactly\footnote{Cf. the discussion in Sec. 3 
of Ref. \cite{Fetter}.}. The second term in (\ref{eq:LDA_high_dens}) is the exchange energy 
$e_{\rm x}$ while the following terms 
correspond to the correlation energy $e_{\rm c}$. For a two-body potential that goes like $1/r$, it 
is quite intuitive 
that interaction terms scale at most like the interparticle spacing, or like 
$\rho^{1/3}$. The different terms of $\frac{E}{N}$ in Eq. (\ref{eq:LDA_high_dens}) are displayed in
Fig. \ref{fig2}, where $\frac{E}{N}$ is written as $e$. The total energy has a minimum that mainly results from the balance between the 
kinetic term that goes like $\rho^{2/3}$ and the exchange term that goes like $\rho^{1/3}$, 
although the correlation energy plays some role. The minimum corresponds to $r_s$ around 3.8, 
surprisingly close to the values for real metals that are in the range 3-5. 
More accurate expressions
for the total energy of the uniform electron gas, that are also valid at lower densities, have
been derived and are available \cite{PW.81}. Below, we shall compare with the nuclear case. 

The so-called Local Density Approximation (LDA) amounts to assuming that, in the vicinity of a given point
$\vec r$, the exchange-correlation energy of any electron system can be approximated
with that of the uniform gas with the same density. In practice, one writes
\begin{equation}
E_{\rm xc} = \int d^3r\ e_{\rm xc}^{\rm hom}[\rho(\vec r)] \rho(\vec r),
\end{equation}
where we have stressed that the exchange-correlation energy has been taken from the homogeneous
gas by means of the superscript. LDA usually produces an overbinding of atoms and molecules,  
that can be corrected by introducing a dependence of $E_{\rm xc}$  on
the gradient of the density, $\nabla\rho$ (Generalized Gradient Approximation, or GGA). 
At present, increasingly sophisticated functionals that depend also on higher derivatives of the
density are discussed and/or start to be built. These various degrees of sophistication are
pictorially referred to as a ``Jacob's ladder'' \cite{PS.01}. As a last rung of the ladder, 
a possible dependence on the unoccupied quantum states of the system is postulated. We shall
come back to these points when discussing the nuclear case.

\section{The nuclear case: the mean-field picture and Hartree-Fock theory}\label{sec:HF}

There are analogies but also important differences between the many-electron problem and the many-nucleon
problem. In the latter case there is no external potential, that is, nuclei are self-bound objects. 
In addition, while in Eq. (\ref{eq:H1}) we have written only a two-body interaction $V(i,j)$,
three-body interactions $V(i,j,k)$ are relevant for nuclei and four-body interactions cannot be
completely ruled out. As compared with the Coulomb force, the nuclear interaction 
is short-ranged (the range is of the order of 1 fm), strongly spin-dependent, and characterised by many terms (central,
spin, spin-orbit, tensor) of similar importance. 

Electron scattering experiments
suggest that ordinary nuclei (except the lightest ones) are characterised
in their inner part
by an approximately constant density, whose value is $\rho_0 \approx$ 0.16 fm$^{-3}$ 
(saturation density) as it has 
been mentioned in the Introduction. 
The nuclear surface, although very important for nuclear properties, 
has a thickness of only $\approx$ 0.6 fm. 
As a consequence, the interparticle distance 
$2r_0$ is $\approx$ 2.4 fm and is larger than the range of the nuclear force. In this
respect, the nucleus is not a very dense system. 
Together with the role played by the Pauli exclusion principle, 
this explains why nucleons do not
experience the mutual interaction so often; in fact, it is known experimentally that the nucleon 
mean free path in nuclei is or the order of, or larger than, the nuclear radius. 

Historically, these evidences have led to assuming the validity of mean-field theory, namely of the
picture that nucleons
move in a one-body potential that results from
the average of the interactions with all other nucleons. It has been deemed to be 
too hard or somehow not so relevant, 
for many years, trying to deduce the mean-field from a complicated 
NN Hamiltonian. Empirical potentials like the well-known Woods-Saxon have been
often taken for granted. Elegant and sophisticated models have been developed by 
the Copenhagen school \cite{BM1,BM2}, or by Landau and Migdal \cite{Migdal}, or by Solovev 
and co-workers \cite{Solovev}, based on single-particle and collective excitations on
top of the mean-field, without daring to ask the question whether this mean-field can
be derived from an underlying theory, or whether binding energies can be calculated therefrom. 
This lecture deals instead with a unified picture in which the {\em effective} mean field can
be derived within the DFT framework.

In the 1970s and 1980s, different authors have started to propose effective Hamiltonians
with the aim of using them within a microscopic mean-field theory, that is,
within Hartree-Fock (HF). One starts from an Hamiltonian analogous to (\ref{eq:H1}),
\begin{equation}
H = \sum_{i=1}^N -\frac{\hbar^2}{2m}\nabla_i^2 + \frac{1}{2}\sum_{i \ne j = 1}^N
V(i,j),
\end{equation}
where the difference between the neutron and proton mass has been neglected for the sake of
simplicity, and defines an energy functional as
\begin{equation}\label{eq:Slaterdet}
E_{\rm HF}[\rho] = \langle \Phi \vert H \vert \Phi \rangle,
\end{equation}
where $\vert \Phi \rangle$ is the most general Slater 
determinant made up with single-particle wave functions $\phi_j$. The minimization of this
functional, with the same constraint as 
in Eq. (\ref{eq:minKS}), namely
\begin{equation}
\frac{\delta}{\delta \rho(\vec r)}\left( E_{\rm HF} - \varepsilon
\int d^3r^\prime\ \phi_j^*(\vec r^\prime) 
\phi_j(\vec r^\prime) \right) = 0,
\end{equation}
can be carried out in practice replacing the variation with respect to
$\rho$ by the variations with respect to $\phi_j^*$, and this  
leads to the well-known HF equations:
\begin{eqnarray}\label{eq:HF}
-\frac{\hbar^2}{2m}\nabla^2\phi_j(\vec r) + \sum_{l=1}^N \int d^3r^\prime\
\phi_l^*(\vec r^\prime) V(\vec r, \vec r^\prime) 
\left( \phi_l(\vec r^\prime) \phi_j({\vec r})
- \phi_l(\vec r) \phi_l(\vec r^\prime) \right) & = & 
\varepsilon_j \phi_j(\vec r),
\nonumber \\
\langle j \vert \frac{-\hbar^2}{2m}\nabla^2 \vert j \rangle + 
\sum_{l=1}^N \langle jl \vert V\left( 1-P_{12} \right) \vert jl \rangle & = & \varepsilon_j,
\end{eqnarray}
where $P_{12}$ exchanges the particles 1 and 2. The total energy reads
\begin{equation}\label{eq:E_HF}
E_{\rm HF} = T + \frac{1}{2} \sum_{jl} \int d^3r\ d^3r^\prime\ \phi^*_j(\vec r)
\phi^*_l(\vec r^\prime) V(\vec r,\vec r^\prime) \left( \phi_j(\vec r)\phi_l(\vec r^\prime) -
\phi_l(\vec r^\prime)\phi_j(\vec r) \right),
\end{equation}
where $T$ has the same form as in Eq. (\ref{eq:Ekin}). We are going now to illustrate that 
a na\"ive picture of the nuclear mean-field is doomed
to fail, even in the simple case of uniform nuclear matter that is the analogous 
system with respect to the electron gas that has been previously discussed. In particular, 
all attempts to account for the empirical evidence(s) using a {\em density-independent} 
effective interaction $V$, at the HF level, have not been successful. 

\section{Uniform nuclear matter}\label{sec:matter}

As we have mentioned in the previous Section, the inner part of nuclei displays 
an approximately constant density, $\rho_0$. In nuclei with $N\approx Z$, proton and neutron 
densities do not differ too much whereas in nuclei with neutron excess there 
is a spill-out of the neutron density but not a large difference in the inside part. 
Thus, we can picture the inside as similar to an extended system having equal number of protons 
and neutrons and equilibrium density equal to $\rho_0$, that is, similar to a piece of 
symmetric nuclear matter (SNM). 
The Bethe-Weisz\"acker formula for the energy per particle $e\equiv\frac{E}{A}$, namely
\begin{equation}
e(A,Z) = a_V - a_S A^{-1/3} - a_A \left( \frac{A-2Z}{A} \right)^2 - a_C \frac{Z^2}{A^{4/3}},
\end{equation}
has an associated volume term $a_V$ of about 
-16 MeV. If we consider uniform SNM all terms of the mass formula vanish except the volume term. 
Therefore, it is customary to assume that the energy per particle of this system, at the 
equilibrium density $\rho_0$, takes the value $e_0$ = -16 MeV. As we have stressed already, 
this is called the saturation point for SNM.

Around this point, which is a minimum in the energy per particle $e$, one 
expand this function as
\begin{equation}
e = e_0 + \frac{1}{2}K_\infty\left( \frac{\rho-\rho_0}{3\rho_0} \right)^2 + \ldots,
\end{equation}
where $K_\infty$ is called nuclear incompressibility, it is proportional to the second derivative
$\textstyle{\frac{d^2e}{d\rho^2}}$ and, as such, it is related to the stiffness of nuclear
matter under compression. Although this quantity can be only indirectly related to observables,
like the properties of the compressional-type vibrations of finite nuclei, some bounds on its values
have been established in the last decades (see \cite{GC.tbp} and references therein). 

One can also
consider the case of asymmetric matter, in which we expect that the energy per
particle must depend both on neutron and proton densities $\rho_n$ and $\rho_p$.
With a simple change of variables, we can use the total density $\rho$ and
the local neutron-proton asymmetry,
$\textstyle {\beta \equiv {\rho_n-\rho_p \over \rho}}$.
By making a Taylor expansion in $\beta$ and retaining only
the quadratic term (odd powers of $\beta$ are obviously forbidden
due to isospin symmetry), we can write
\begin{equation}\label{def_sym}
e(\rho,\beta) = e(\rho,\beta=0) +
S(\rho) \beta^2,
\end{equation}
where the first term on the r.h.s. is the energy per particle of SNM that we have so far discussed, while
the second term defines
the so-called symmetry energy $S(\rho)$. From the latter equation, the symmetry energy can be
easily understood as the energy per particle that is needed to change symmetric matter into neutron matter.

The symmetry energy can, in turn, be expanded as a function of
$\rho$ and, as the saturation point of SNM is a useful reference, one usually writes
\begin{equation}
S(\rho) = J + L \left( \frac{\rho - \rho_0}{3\rho_0} \right)
+ \frac{1}{2} K_{sym} \left( \frac{\rho - \rho_0}{3\rho_0} \right)^2 + \ldots,
\end{equation}
where $J$ is the value of the symmetry energy at saturation density and the other parameters are related
to the first and second derivatives at the same point. In particular, $L$ is often referred to as the
``slope parameter''. Constraints on the values of $J$ and $L$ have been established
by the study of phenomena in which a neutron-proton imbalance is created at different densities, like
oscillations where protons and neutrons vibrate out of phase or heavy-ion collisions where projectile
and target have different composition; masses as a function of the neutron excess, or observations
of neutron stars, can also help in constraining the values of $J$ and $L$
(cf. Refs. \cite{topical_volume,Lattimer:2013,Li:2013,oertel2017,Horowitz:2014}).

\section{Failure of mean-field with simple forces and the need for DFT}\label{sec:MF}

In SNM the Fermi momentum is
given by\footnote{The presence of two neutrons and two protons for each value of the
momentum $\vec k$, namely the degeneracy $g$ = 4, makes the factor in the next equation 
different from that of the electronic case.}
\begin{equation}\label{eq:kF}
k_F = \left( \frac{3\pi^2\rho}{2} \right)^{1/3},
\end{equation}
and, at saturation density, it takes the value $\approx$ 1.33 fm$^{-1}$.
The wave functions in a uniform system are plane waves due to translational
invariance. This simplifies the calculation of the total energy (\ref{eq:E_HF}),
in keeping with the well-known replacements
\begin{eqnarray}\label{eq:PW} 
\phi_j(\vec r) & \rightarrow & \frac{1}{\sqrt{\Omega}}e^{i{\vec k}\cdot{\vec r}} \nonumber \\
\sum_j & \rightarrow & \frac{g\Omega}{(2\pi)^3}\int\ d^3k \ , 
\end{eqnarray}
where $g$ = 4 is the degeneracy and $\Omega$ is the quantisation volume.
The kinetic part $T$ of the total energy can be calculated straighforwardly and re-expressed in terms
of the density thanks to Eq. (\ref{eq:kF}). The kinetic energy per particle, $t$, is  
\begin{equation}
t = \frac{T}{A} = \frac{2}{\pi^2}\frac{\hbar^2}{2m}\frac{k_F^5}{5}\frac{\Omega}{A} = 
\frac{3}{5}\left( \frac{3\pi^2}{2} \right)^{2/3}
\frac{\hbar^2}{2m}\rho^{2/3}.
\end{equation}

Now, the question arises under which conditions, within simple HF, an effective force $V$ allows nuclear saturation. 
In terms of radial dependence, Yukawa functions may constitute a choice as they resemble what
is deduced from the Fourier transform of a massive boson propagator. Gaussian functions
have also been, and still are, used because they allow an easy calculation of the matrix elements 
on a harmonic oscillator basis \cite{Gogny:1975}. From the viewpoint of EFT, if one is interested 
in nuclear properties at the scale of tens of MeV or less, the details of the radial shape of the 
force on a scale around 1 fm should be unimportant\footnote{The reader should remember that
$\hbar c$ = 197.3 MeV fm, and this allows making the equivalence between 1 fm and about 200 MeV. Natural 
units embed such concepts but will not be used throughout this lecture.}. 
Intuitively, nuclear saturation should emerge from a balance between some attractive and repulsive 
mechanisms and we will discuss briefly the minimal number of terms 
that have to be introduced to this aim, 
sticking to a nonrelativistic picture as already mentioned\footnote{The mechanism for saturation 
in a covariant theory (balance between scalar and vector potentials) is discussed in the
original paper by Serot and Walecka \cite{SerotWalecka}, as well as in the aforementioned references
\cite{LNP.641,VALR.05}.}.

\begin{figure}[h!]
\centering\includegraphics[width=0.7\textwidth]{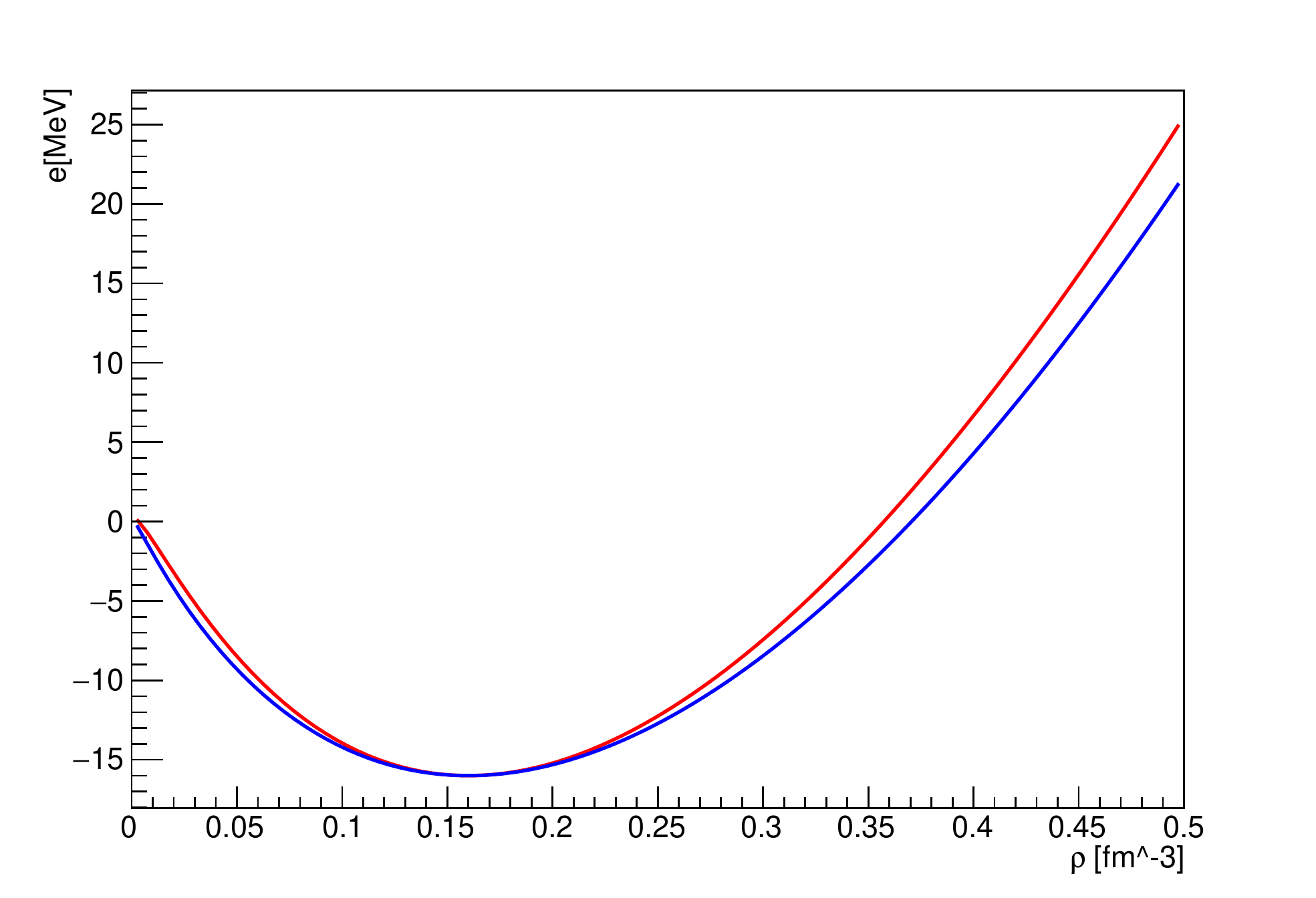}
\caption{Energy per particle in symmetric nuclear matter, calculated either with a force of the type
(\ref{eq:BB}) and parameters $S$ = -429.83 MeV, $m$ = 1.3218, $\mu$ = 0.8343 fm (blue curve), or
with a simplified Skyrme functional of the type (\ref{eq:simpleSkyrme}) and parameters $t_0$ = 
-2552.84 MeV$\cdot$fm$^3$, $t_3$ = 16694.7 MeV$\cdot$fm$^{3(\alpha+1)}$, $\alpha$ = 0.20309.}
\label{fig3}
\end{figure}

For a single Gaussian interaction having a range $\mu$,
\begin{equation}\label{eq:singleGaussian}
V(\vec r_1, \vec r_2) = Se^{\textstyle -\frac{\vert \vec r_1 - \vec r_2\vert^2}{\mu^2}},
\end{equation}
the direct and exchange terms of the total energy per particle, $v_H$ and 
$v_x$, calculated from Eq. (\ref{eq:E_HF}), 
read\footnote{The calculation is straightforward, although some integrals 
are not elementary as discussed in the
Coulomb case in Ref. \cite{Fetter}.}
\begin{eqnarray}\label{eq:E_HF_Gaussian}
v_H & = & \frac{S}{2}\mu^3\pi^{3/2}\rho, \nonumber \\
v_x & = & -\frac{S}{2}\frac{1}{\sqrt{\pi}}g(\mu k_F),
\end{eqnarray}
where $g(x) = \frac{2}{x^3}-\frac{3}{x}-\left( \frac{2}{x^3}-\frac{1}{x}\right) e^{-x^2}+
\sqrt{\pi}{\rm erf}(x)$ \cite{Sel.14,Accorto}. One can immediately notice that the direct 
term goes like $\rho$, as it must be for a short-range two-body interaction, and in contrast with the
case of the electron gas. The exchange term has opposite sign, and a more subtle
density dependence that can nevertheless be easily inferred. 
The force is attractive
($S<0$), and it turns out by inspection that the repulsion from the exchange term plus the kinetic energy is far
too weak to allow saturation. A possible way out, as already suggested long ago by 
Brink and Boeker \cite{BB.67}, is to introduce exchange terms in the interaction. For instance,
Eq. (\ref{eq:singleGaussian}) can be generalised to
\begin{equation}\label{eq:BB}
V(\vec r_1, \vec r_2) = S(1-m+mP_M)e^{\textstyle -\frac{\vert \vec r_1 - \vec r_2\vert^2}{\mu^2}},
\end{equation}
where $P_M$ is the operator that exchanges the particles 1 and 2. 
Then, the potential energies per particle of Eq. (\ref{eq:E_HF_Gaussian}) become
\begin{eqnarray}\label{eq:E_HF_BB}
v_H & = & \frac{S}{8}\left( 4 - 5m \right) \mu^3\pi^{3/2}\rho, \nonumber \\
v_x & = & \frac{S}{2} \left( 5m - 1 \right) \frac{1}{\sqrt{\pi}}g(\mu k_F).
\end{eqnarray}
In this way, the weight of the different terms is changed, and saturation becomes possible for values of 
$\mu$ that lie approximately in the range 0.5 fm $< \ \mu\ <$ 1.5 fm \cite{BB.67}. 
An example of calculation of the HF energy per particle, performed with a force of this 
type \cite{Accorto}, is shown in Fig. \ref{fig3} and compared with the result of a Skyrme
functional that is discussed in the next Section. Similar arguments apply in the case of a Yukawa 
interaction: only exchange terms
allow saturation, as it was remarked in Sec. 2.3.1 of \cite{Blaizot1980} where one can also find
the formulas that are analogous to Eq. (\ref{eq:E_HF_BB}) (cf. also \cite{Rancati}).

The simple forces that we have described earlier in this Section do not possess necessarily enough flexibility
to reproduce the empirical values of $K_\infty$, $J$ and $L$. Certainly, one could complicate
them and introduce more Gaussian (or Yukawa) terms with different ranges, different exchange operators etc. 
What seems, so far, impossible to obtain in such na\"ive mean-field scheme
is the empirical value of the effective mass, $m^*/m$. 

In a uniform system, the most general way
to write the single-particle dispersion relation, namely the relation between energy and momentum (which is
the only available quantum number), reads
\begin{equation}\label{eq:epsilon1}
\varepsilon = \frac{\hbar^2 k^2}{2m} + \Sigma(k,\varepsilon) \equiv \frac{\hbar^2 k^2}{2m^*},
\end{equation}
where $\Sigma$ is the single-particle self-energy. 
The second equality is the definition of the effective mass, 
and by further elementary steps \cite{MBBD.85} one can arrive
at
\begin{equation}
\frac{m^*}{m} = \left( 1 + \frac{m}{\hbar^2 k}
\frac{\partial\Sigma}{\partial k} \right) \left( 1 - \frac{\partial\Sigma}{\partial\varepsilon} 
\right)^{-1}.
\end{equation}
The HF equations of the second line in Eq. (\ref{eq:HF}), in the case in which the wave function 
is a plane wave as in (\ref{eq:PW}), can be easily cast in the form (\ref{eq:epsilon1}). In a simple way, 
one obtains
\begin{equation}
\Sigma_{\sigma\tau}(k,\varepsilon) = \sum_{\sigma^\prime\tau^\prime} \frac{\Omega}{(2\pi)^3} 
\int d^3k^\prime\ 
\langle \vec k,\sigma,\tau; \vec k^\prime,\sigma^\prime,\tau^\prime \vert 
V\left( 1-P_{12} \right) \vert \vec k,\sigma,\tau; \vec k^\prime,\sigma^\prime,\tau^\prime \rangle, 
\end{equation}
where we have explicitly introduced the sum over the spin and isospin degrees of freedom 
($\tau$ labels either neutrons or protons). 
We expect that $\Sigma$ depends on $\sigma$ ($\tau$) only in
the case of polarized (neutron-proton asymmetric) matter. Once $V$ is given, $\Sigma$ can be
calculated and the effective mass can be extracted therefrom. One can explicitly see that, 
within HF, $\Sigma$ does not depend on $\varepsilon$. The non-trivial value of $m^*/m$ is brought by
the $k$-dependence of $\Sigma$ that comes from exchange terms. 

From its definition (\ref{eq:epsilon1}), the effective mass is clearly related to the level density: the larger
its value, the smaller the level spacing. In a non-uniform system the picture is less simple, 
but nevertheless the whole nuclear phenomenology is consistent with a value of the effective mass
$m^*/m$ lying in the range $\approx$ 0.7-1. This has been established for quite a long time 
(cf. the very comprehensive review paper \cite{MBBD.85}, where also the density and energy
dependence of the effective mass is addressed). The simple HF that we have described so far, leads
to much smaller values (between 0.2 and 0.4). This problem is discussed in detail 
in Refs. \cite{Weisskopf,Nakatsukasa:2016,1712.03003} (cf. also \cite{Accorto,Rancati,Romano}).

So far, it has {\em not} been possible to design an
effective $V$ that provides a successful description both of the bulk nuclear properties 
(nuclear saturation and properties of uniform matter around $\rho_0$, as well as masses and
radii of finite nuclei as we discuss below) and of nuclear spectroscopic properties (effective
mass, {\em viz.} level density), without introducing a {\em density dependence} in the Hamiltonian.

A density-dependent $V[\rho]$ or $H[\rho]$ has, generally speaking, 
conceptual problems unless one considers it merely as 
a way to generate an energy functional through
Eq. (\ref{eq:Slaterdet}), that is,
\begin{equation}\label{eq:dd}
E[\rho] = \langle \Phi \vert H[\rho] \vert \Phi \rangle.
\end{equation}
In this respect, we can conclude that DFT, and not HF, is a viable theory for nuclei in our
current understanding. A similar argument can be found under different forms in the literature 
\cite{Nakatsukasa:2016}.

\section{Examples of nuclear EDFs}\label{sec:examples}

The finite-range Gogny force \cite{De.80,BGG.91} is the generalisation of the Brink-Boeker force
that we have introduced in the previous Section. It reads:
\begin{eqnarray}
V_{\rm Gogny}(\vec r_1, \vec r_2) & = & \sum_{j=1}^2 e^{\textstyle \frac{\vert \vec r_1-\vec r_2\vert^2}
{\mu_j^2}}\left( W_j + B_j P_\sigma - H_j P_\tau - M_j P_\sigma P_\tau \right) \nonumber \\
& + & t_3\left( 1 + x_0 P_\sigma \right) \delta(\vec r_1-\vec r_2) \rho^\alpha\left(
\frac{\vec r_1 + \vec r_2}{2} \right) + iW_{ls}\left( \vec \sigma_1 + \vec \sigma_2 \right) \cdot
\vec k^\dagger \times \delta(\vec r_1-\vec r_2) \vec k,
\end{eqnarray}
where $P_\sigma=\frac{1}{2}(1+\vec\sigma_1\cdot\vec\sigma_2)$ is the spin-exchange operator, 
$P_\tau=\frac{1}{2}(1+\vec\tau_1\cdot\vec\tau_2)$ is the isospin-exchange operator, 
$\vec k=-\frac{i}{2}\left( \vec \nabla_1 - \vec \nabla_2 \right)$ is the relative momentum operator
acting at right and $\vec k^\dagger$ is the adjoint operator acting at left. The interaction is
the sum of two Gaussians with exchange operators, a density-dependent term and a spin-orbit term.
The density dependent term has been deemed to be essential to obtain a reasonable single-particle 
level density (cf. our discussion of the effective mass in the last Section). This term must be zero-range to avoid
ambiguities on the point where the density must be evaluated. The spin-orbit term is also zero-range, 
for simplicity. This interaction has 14 free parameters to be adjusted.

Another class of successful effective interactions is based on the zero-range, momentum-dependent
Skyrme ansatz \cite{VB.72,Cha.97,Cha.98}:
\begin{eqnarray}\label{eq:VSkyrme}
V_{\rm Skyrme}(\vec r_1, \vec r_2) & = & t_0 \left( 1+x_0 P_\sigma \right) \delta(\vec r_1 - \vec
r_2) + \frac{1}{2}t_1 \left( 1+x_1 P_\sigma \right) \left( \vec k^{\dagger 2}\delta(\vec r_1 - \vec 
r_2) + \delta(\vec r_1 - \vec r_2)\vec k^2 \right) \nonumber \\
& + & t_2 \left( 1+x_2 P_\sigma \right)\vec k^\dagger \cdot \delta(\vec r_1 - \vec r_2) 
\vec k + \frac{1}{6}t_3 \left( 1+x_3 P_\sigma \right) \delta(\vec r_1 - \vec r_2) 
\rho^\alpha\left( \frac{\vec r_1 + \vec r_2}{2} \right) 
\nonumber \\
& + & iW_{0}\left( \vec \sigma_1 + \vec \sigma_2 \right) \cdot
\vec k^\dagger \times \delta(\vec r_1-\vec r_2) \vec k,
\end{eqnarray}
where now there are 10 free parameters to be adjusted. As we have just mentioned, 
a density-dependent force is just a way to generate an energy functional through 
Eq. (\ref{eq:dd}). 
It is a good exercise for the reader to show that 
a simplified Skyrme force without momentum and spin dependence,
\begin{equation}\label{eq:simpleSkyrme}
V(\vec r_1, \vec r_2) = t_0 \delta(\vec r_1 - \vec r_2) + 
\frac{1}{6}t_3 \left( 1+x_3 P_\sigma \right) \delta(\vec r_1 - \vec r_2)
\rho^\alpha\left( \frac{\vec r_1 + \vec r_2}{2} \right),
\end{equation}
generates for even-even systems the functional
\begin{eqnarray}\label{eq:simplestEDF}
E[\rho_n,\rho_p] & = & \int d^3r\ {\cal E}[\rho_n,\rho_p], \nonumber \\
{\cal E}[\rho_n,\rho_p] & = & \frac{\hbar^2}{2m}\tau + \frac{1}{2}t_0 \left[ \rho^2 - \frac{1}{2} \left( 
\rho_n^2 + \rho_p^2 \right) \right]
+ \frac{1}{12}t_3 \left[ \rho^{\alpha+2} - \frac{1}{2} \rho^\alpha \left(
\rho_n^2 + \rho_p^2 \right) \right].
\end{eqnarray}
In these latter equations, the total energy $E$ is written in terms of the energy density $\cal E$. 
Thus, one speaks (in this and all cases to be discussed below) of $\cal E$ as of an 
energy density functional (EDF), namely an energy density that depends on functions like 
$\rho_n$ and $\rho_p$. Zero-range forces generate local functionals (at variance with the Coulomb 
case that has been previously discussed). Fractional power dependences like $\rho^\alpha$ cannot 
come from density-independent two-body, three-body, or many-body forces. In other terms, they
mimic many-body effects in a more subtle way. It has been known for many years now that only
functionals in which such fractional powers appear, reproduce the empirical values of $K_\infty$
\cite{Blaizot1980,GC.tbp}. The kinetic part of the Hamiltonian introduces the first term 
appearing in the r.h.s. of Eq. (\ref{eq:simplestEDF}), that depends on the so-called kinetic energy density,
\begin{equation}
\tau = \sum_j \vert \vec \nabla \phi_j \vert^2.
\end{equation} 
The whole Skyrme force (\ref{eq:VSkyrme}) 
includes momentum-dependent terms that bring a dependence on $\tau$ into the 
potential energy density as well, together with a dependence on $\vec \nabla \rho$. The complete EDF 
associated with a Skyrme force can be found, for even-even systems, in Refs. \cite{Cha.97,Cha.98}. 
More densities appear like the spin-orbit densities $\vec J(\vec r)$. Interestingly, terms that 
depend on the same kinds of densities that have been introduced for electron
systems in \cite{PS.01}, characterise the Skyrme EDFs. A finite-range force like the Gogny interaction
produces a non-local EDF.

In fact, the very idea that an Hamiltonian including a density-dependent force is {\em the} tool to
generate an EDF has been abandoned by most groups. A functional can be directly parameterized 
in terms of local densities, without any reference to an underlying interaction. A pioneering step 
in this direction was taken by Reinhard and Flocard \cite{RF.95} more than twenty years ago, as they
wrote the spin-orbit part of their EDFs without reference to a specific form of the force. Later,
this has become the customary procedure, e.g. for functionals of the UNEDF family \cite{Kortelainen2014}.

All possible local densities that can appear in an EDF, have been classified in 
Refs. \cite{Engel:1975,DD.95,DD.96,BHF.03} (see also \cite{DD.er}). The nuclear EDF must be invariant 
with respect to parity, time-reversal, rotational, translational and isospin transformations (details 
can be found in the quoted works). Time-reversal plays a special role here. Densities can be either 
time-even or time-odd but the latter vanish in even-even systems. Thus, the EDF can be made 
up with terms that are bilinear in either types of densities, but the terms that are bilinear in 
time-odd densities do not vanish only in systems with an odd number of particles. A current open questions for practitioners and
not only, is to which extent higher and higher gradients of the density are needed for an EDF
in order to satisfactorily account for nuclear properties \cite{Raimondi2011,Becker1,Becker2}. 

This discussion is of course not exhaustive. Other forces have been considered as generators of
EDFs, for examples semi-realistic forces of Yukawa type with density-dependent terms \cite{Nakada2008}.
Other kinds of functionals have been proposed by Fayans and collaborators \cite{FTZ.94,Fayans1998}. 
We do not discuss all the attempts to derive functionals from underlying theories like 
Br\"uckner-Hartree-Fock \cite{BCPM} or chiral forces.

\section{Examples of calculations of ground-state properties}\label{sec:examples_gs}

The main observable that one aims at calculating with an EDF, is the total energy which is
actually the binding energy of a nucleus and can be compared with very accurate experimental
data. The typical errors 
are of the order of $\approx$ 1-2 MeV.
While this accuracy may be thought to be small, at least in comparison with total binding energies
that span values like 10$^2$-10$^3$ MeV, one should keep in mind that nuclear processes like reactions and
decays depend on differences of binding energies, that may be small. This is the motivation for
reducing these errors with respect to experiment. Models that include macroscopic ingredients
in alternative, or in addition, to the pure implementation of DFT may reach accuracies of the order
of hundreds of keV (cf. the review paper \cite{Lunney} or, for more recent advances, 
\cite{Goriely} and references therein). 

\begin{figure}[t!]
\centering\includegraphics[width=0.8\textwidth]{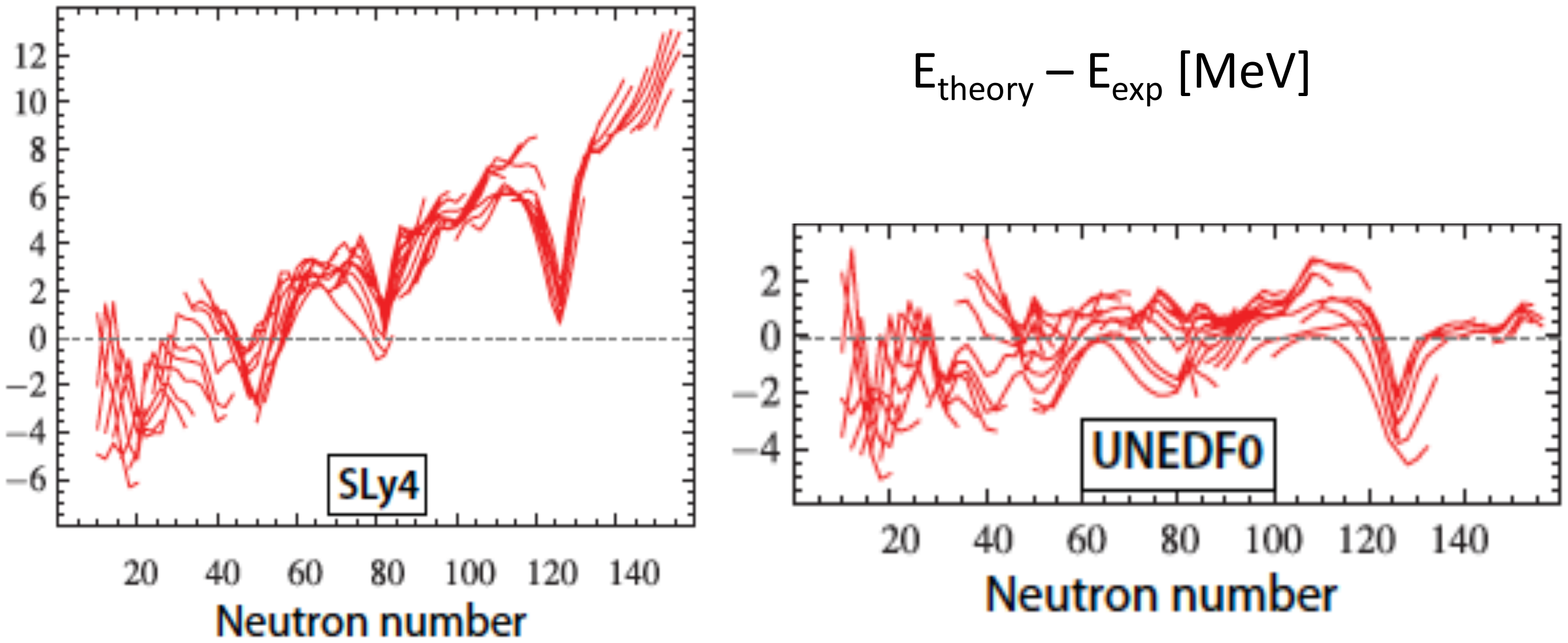}
\caption{Comparison between theoretical and experimental binding energies. Calculations are
performed with different EDFs, SLy4 from Ref. \cite{Cha.98} and UNEDF0 from Ref. 
\cite{Kortelainen:2010}. Figure taken and adapted from Ref. \cite{Kortelainen:2010}.}
\label{fig_masses}
\end{figure}

In Fig. \ref{fig_masses} we show examples of comparison between experimental binding energies and
the result of DFT calculations. The Skyrme functional SLy4 \cite{Cha.98} has been fitted by
using masses and charge radii of basically only magic nuclei. This is probably the reason of the
archlike behaviour, which is common to many EDFs. Reproducing with similar accuracy closed-shell
and open-shell nuclei still represents a challenge. More modern functionals like UNEDF0
\cite{Kortelainen:2010} seem to behave much better than SLy4, although the trend of the error as a 
function of the neutron number is not really
flat. The reader should be also aware of the significant recent progress in DFT calculations using 
covariant functionals, not discussed in this contribution: a careful analysis of the mass residuals, and
their link(s) with other features of the EDFs, is carried out in Ref. \cite{rel_masses}. 
To which extent some correlations that are relevant for nuclear masses cannot be captured by
DFT calculations, is still an open question.

EDFs have been applied to predict the limits of nuclear stability, namely the position of the
drip lines that we have mentioned in the Introduction \cite{Erler2012,Afanasjev2013}. Other relevant
ground-state observables are radii and density distributions. Charge radii are available for 
comparison with experiment: EDFs may display errors of the order of 0.02-0.03 fm, typically.
As far as isotopic trends of radii are concerned, while in many instances EDFs perform well,
there are still specific unresolved issues (cf. Fig. 11 of \cite{BHF.03}: the isotopic dependence of 
the charge radii along the Ca chain is a typical case where EDFs fail). Last but not least,
there is strong current interest in measuring neutron radii or density distributions; EDFs 
provide predictions than span a relatively broad interval for neutron radii, and fixing this
observable would be of paramount importance for our understanding of the nuclear symmetry energy (see 
the discussion in Sec. \ref{sec:matter}, and the references therein). 

\section{Intrinsic density}\label{sec:intrinsic}

The biggest difference between DFT in finite electronic systems and nuclei consists in
the fact that in the former case, in most instances, the fixed ion positions constrain the shape
of the system in the laboratory frame. In nuclei, that are self-bound systems, this is not the case. 
The usual HK theorem, as it has been argued by several authors \cite{Engel:2007}, is irrelevant
to the nuclear case because it concerns the laboratory density, while experiments probe the
intrinsic density (relative to the nuclear center of mass). 

Nonetheless, it has been proven that,
at least in principle, given an arbitrary Hermitian operator $\hat Q$, one can build
an energy functional depending on $Q(\vec r) \equiv \langle \hat Q(\vec r) \rangle$ that is universal in the HK sense
and has its minimum at the correct value of $Q$ with the correct energy. In this respect,
one can replace the laboratory density with the intrinsic density in the HK theorem
\cite{Engel:2007,Barnea:2007,MBS:2009}. 

A very interesting point is that the intrinisic density appears to break, in many cases, the symmetries
associated with the Hamiltonian as we discuss in the next Section. 

\section{Symmetry breaking and restoration}\label{sec:sym}

In general, if the Hamiltonian of a system commutes with a given symmetry operator $S$,
\begin{equation}
\left[ H, S \right] = 0,
\end{equation}
one expects to be able to find eigenstates of the system which are also eigenstates of $S$. 
The problem with this statement of principle comes when one is forced to treat strongly
correlated systems and to resort to approximations. In that case, a symmetry-conserving solution
may be much less realistic, within the approximated framework, than a symmetry-breaking one.

The first example is that of translational symmetry. A very simple example borrowed 
from \cite{Goutte,Lacroix} will highlight the above statements. Let us consider two particles
in 1D, confined in the interval $-L/2 < x < L/2$, interacting through an harmonic potential,
\begin{equation}\label{eq:H1D}
H = \frac{p_1^2}{2m} + \frac{p_2^2}{2m} + \frac{C}{L^2}\left( x_1 - x_2 \right)^2.
\end{equation}
By using the center-of-mass and relative coordinates
$R={\textstyle \frac{x_1+x_2}{2}}$ and $r=x_1-x_2$, together with the associated momenta $P$ and $p$, 
respectively, this Hamiltonian can be separated into center-of-mass and intrinsic Hamiltonians,
\begin{equation}
H = \frac{P^2}{4m}+\frac{p^2}{m}+\frac{C}{L^2}r^2 = H_{COM} + H_{intr}.
\end{equation}
This Hamiltonian clearly commutes with the total momentum $P$, that is, it manifests the translational
invariance. The exact wave function is known, namely it is a product of a free particle times a harmonic 
oscillator wave function, it is an eigenstate of $P$, and the associated energy is
\begin{equation}\label{eq:exact_twop_1D}
E_{exact} = \frac{P^2}{4m} + \sqrt{\frac{\hbar^2 C}{mL^2}}.
\end{equation}
Let us now imagine that we restrict ourselves to products of independent wave functions. If the two particles
are bosons, or spin 1/2 fermions in a spin-antisymmetric state\footnote{We would like to avoid 
antisimmetrization in order to simplify the pedagogical argument we are developing.}, the total wave function 
is 
\begin{equation}\label{eq:simpleSlater}
\Phi = \phi_1(x_1)\phi_2(x_2).
\end{equation}
Imposing translational invariance here, would amount to take the $\phi$ as plane waves which means  
$\textstyle{\phi_i(x_i)=\frac{exp(ip_ix_i/\hbar)}{\sqrt{L}}}$. 
The expectation value,
\begin{equation}\label{eq:simpleE}
E = \langle \Phi \vert H \vert \Phi \rangle,
\end{equation}
can be easily calculated by using the Hamiltonian in the form (\ref{eq:H1D}), and one can verify that
its value is significantly different from the exact one of 
(\ref{eq:exact_twop_1D}). If we use the 
independent particle ansatz (\ref{eq:simpleSlater}) and we give up the requirement of translational
invariance, the minimization of (\ref{eq:simpleE}) produces instead an intrinisic energy given by
\begin{equation}\label{eq:ESB}
E_{symm.\ breaking} = 2\sqrt{\frac{\hbar^2C}{2mL^2}},
\end{equation}
with an associated wave function that is the product of two localized harmonic oscillators. 
This energy is relatively close to the intrinsic part of the exact one, that is, the second term 
at the r.h.s. of Eq. (\ref{eq:exact_twop_1D}).

The above example should make clear why the choice of using the variational ansatz with
symmetry-breaking wave functions might be preferable to other options. Translational symmetry 
and its breaking characterises all systems, and it is not of special relevance for
the atomic nucleus. On the other hand, rotation and gauge symmetries, that we shall now
discuss, are instead associated with 
the quadrupole and pairing correlations that have been identified as the most important 
correlations in finite nuclei since several decades. 

To make the introduction to rotational symmetry and its breaking as simple as possible, 
let us pick up another very simple example, by considering several particles 
in a 3D harmonic one-body potential plus a zero-range two-body residual interaction. 
The corresponding Hamiltonian reads
\begin{equation}\label{eq:Hqq}
H = \sum_i \frac{{\vec p_i}^2}{2m} + \sum_i \frac{1}{2}m\omega^2r_i^2 
+ \frac{1}{2}\sum_{ij} \epsilon \delta(\vec r_i - \vec r_j),
\end{equation}
where $\epsilon$ is small so that the residual interaction term will be treated as a perturbation of the 
3D harmonic oscillator. The particles in this potential
have eigenstates $\vert nlm\rangle$, with  
associated energies $E_{nl}=\hbar\omega\left( 2n+l-\frac{1}{2}\right)$ and wave functions 
$\phi_{nlm}(\vec r) = \frac{u_{nl}(r)}{r}Y_{lm}(\theta,\phi)$, where the radial part can be explicitly
written in terms of Laguerre polynomials. 
If these particles are fermions, they occupy the states 1s, 1p etc. 

If we have a set of orbitals that are filled and we add one particle, this will occupy the lowest 
unoccupied orbital. The perturbing 
zero-range interaction [last term in Eq. (\ref{eq:Hqq})] has been assumed to be weak and, consequently, 
unable to alter the structure of the orbitals and shells. Nevertheless, if we add a second particle
this will produce at lowest order an energy shift $\Delta E$ given by
\begin{equation}
\Delta E = \epsilon \langle nlm, nlm' \vert \delta(\vec r_1-\vec r_2) \vert nlm, nlm' \rangle,
\end{equation}
where $m,m'$ are the quantum numbers of the two particles in the degenerate levels. 
This shift can be calculated\footnote{The multipole expansion of the delta function, and
the matrix elements of the spherical harmonics, are needed.} and the result reads
\begin{equation}
\Delta E = \epsilon \int dr \frac{u^4_{nl}(r)}{r^2} \sum_\lambda 
\frac{2\lambda+1}{4\pi}\langle lm \lambda 0 \vert lm \rangle 
\langle lm' \lambda 0 \vert lm' \rangle \langle l0 \lambda 0 \vert l0 \rangle^2,
\end{equation}
where the sum is limited by the angular momentum selection rules, that are implicit in the
Clebsch-Gordan coefficients. 

\begin{table}[t]
\caption{Values of $f(m,m')$ defined in the main text.}\label{table1}
\begin{tabular}{rccccc}
\hline
$m/m'$ & -2  &  -1    &  0     &  1     &  2     \\
\hline
 -2 &        &  0.227 &  0.227 &  0.227 &  0.455 \\
 -1 &  0.227 &        &  0.227 &  0.455 &  0.227 \\
  0 &  0.227 &  0.227 &        &  0.227 &  0.227 \\
  1 &  0.227 &  0.455 &  0.227 &        &  0.227 \\
  2 &  0.455 &  0.227 &  0.227 &  0.227 &        \\
\hline
\end{tabular}
\end{table}

The values of of $f(m,m')=\frac{\Delta E}{\epsilon 
\int dr u^4_{nl}(r) r^{-2}}$ are displayed in Table \ref{table1} for the case of $l=2$ 
(d orbitals). 
The diagonal values are excluded from the Table because the two particles cannot occupy the
same quantum states. The largest values are those with $m'=-m$. This can be intuitively understood 
in a qualitative way, since the two orbitals with the same absolute value of $m$ have the largest
spatial overlap so that a short-range interaction has the largest effect. If such short-range
interaction is attractive, two particles will let the system gain energy if the orbitals 
with $\vert m \vert = 2$ are occupied, that is, the system has a non-spherical shape.
This deformation is enhanced by polarization effects, namely by the interaction between the 
particles in the last orbitals and the others. The details of polarization effects go beyond 
the scope of this lecture. Nevertheless, these effects contribute to make the nucleus
deformed as a whole.

Quadrupole deformations are, by far, those 
dominant along the nuclear chart. Nuclei with quadrupole deformation have been identified 
by their rotational bands, that is, in keeping with the fact that their spectra correspond
to those of a rotor \cite{BM2}. Also, octupole correlations and octupole deformations
have been subject of recent interest and have been now unambiguously 
proven \cite{BN.96,Gaffney2013}. 

If the quadrupole deformed configuration of
the system corresponds to a deep minimum of the total energy, that is, the total energy has a 
stiff behaviour around this minimum, DFT describes well, as a rule, the rotational spectra
or at least the lowest sector of the rotational bands \cite{BHF.03}. Phenomena that take place
when the rotation is fast as the angular momentum is large, or hyperdeformation, are 
discussed in the literature and also dealt with in the nuclear DFT context (see e.g. 
Ref. \cite{DD.95}). 

\begin{figure}[t!]
\centering\includegraphics[width=0.4\textwidth]{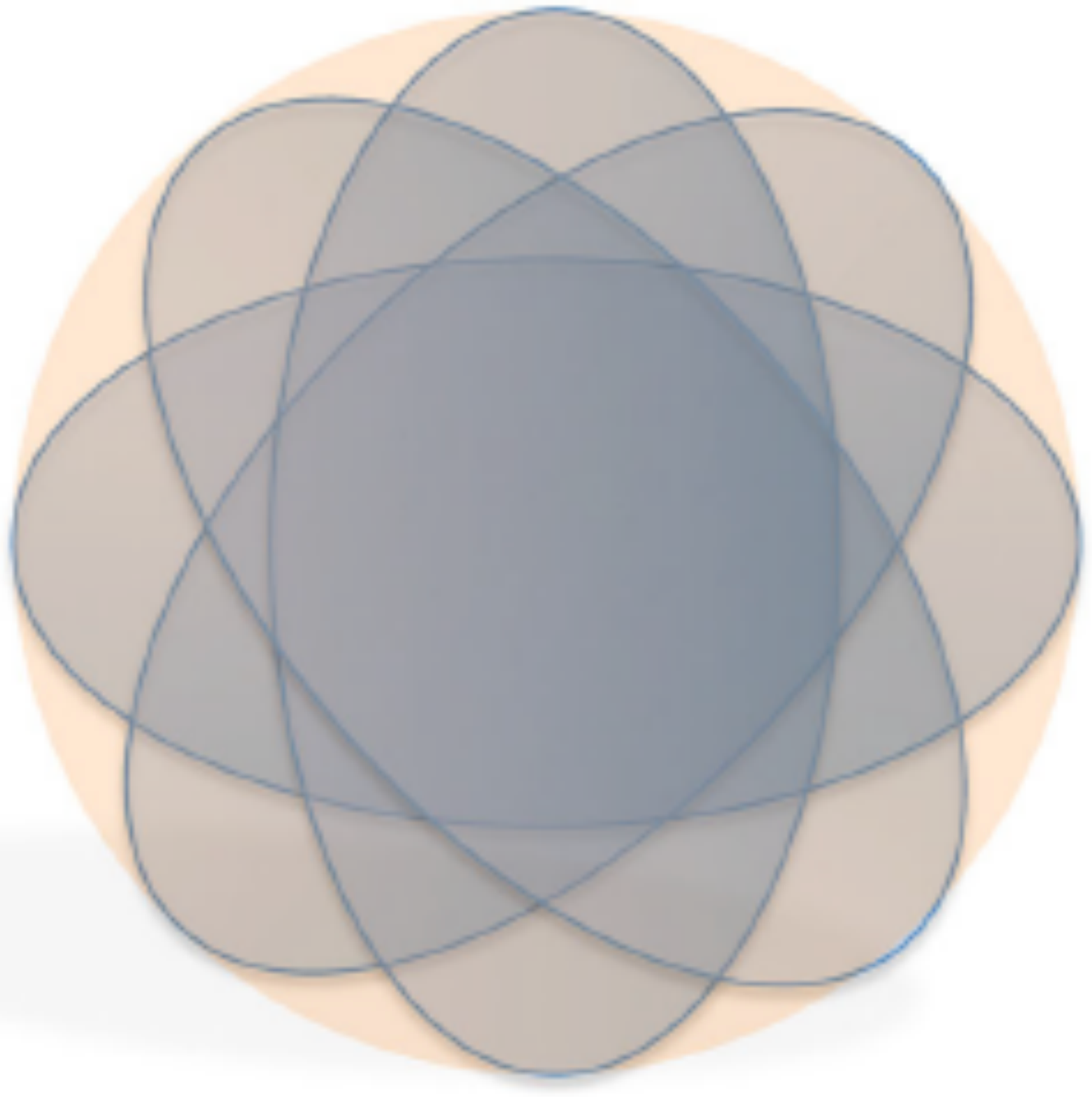}
\caption{Schematic view (in two dimensions) of the deformed nucleus in the intrinsic frame that
rotates in the laboratory frame. See the main text for a discussion. Figure 
adapted from Ref. \cite{Nakatsukasa:2016}.}
\label{fig4}
\end{figure}

As is clear from the previous discussion, deformation occurs in the center-of-mass frame.
Thus, we have spontaneous symmetry breaking {\em in the intrinsic system}. The way in which 
the symmetry is restored is by means of rotation in the laboratory system, that is, 
by superimposing different shapes that are produced through a rotation by a set of Euler angles
$\Omega$. This is depicted schematically, in 2D, in Fig. \ref{fig4}.  
Since the result of the rotation of a spherical harmonic $Y_{lm}$ by $\Omega$ is given by 
$\sum_{m'}{\cal D}^{l}_{m'm}(\Omega)Y_{lm'}$, where ${\cal D}$ is a Wigner function, the wave function associated
with the superposition of states $l,m'$ that are rotated reads
\begin{equation}\label{eq:ang_proj}
\vert nlm \rangle = {\cal N} \int d\Omega\ {\cal D}^{l}_{m'm}(\Omega) \vert nlm'\rangle,
\end{equation}
where ${\cal N}$ is a normalization factor. This can be also interpreted as a projection on 
good angular momentum in the laboratory frame. 
Superimposing different shapes, and restoring the rotational symmetry by means of this
angular momentum projection, is called multi-reference DFT (MR-DFT) in the context of nuclear
EDFs. There is rapid development in recent years concerning MR-DFT, using the Skyrme and
Gogny functionals that have been discussed in Sec. \ref{sec:examples} as well as using the
covariant functionals. 

A similar, yet more abstract, reasoning holds in the case of the pairing correlations \cite{BB.05}. 
Open-shell nuclei are characterised by the fact that a fraction of nucleons around the Fermi 
energy display a superfluid character. Their phenomenology is, as a rule, well
described by the BCS (Bardeen-Cooper-Schrieffer) theory. Within this framework, nucleons
in time-reversal states form so-called Cooper pairs, and the superfluidity is associated 
with the coherent behaviour
of these fermion pairs that are quasi-bosons. A very economic way to introduce pairs
that are made up with fermions in time-reversal states is the BCS ansatz for the many-body
wave function, that reads
\begin{equation}\label{eq:BCS_wf}
\vert \Phi \rangle = \Pi_j \left( u_j + v_j a^\dagger_j a^\dagger_{\tilde j} \right) \vert - \rangle, 
\end{equation}
where $j$ labels a set of single-particle orbitals like the $\phi_j$ that have been previously
discussed, $a_j$ and $a^\dagger_j$ are, respectively, the annihilation and creation operators
that destroy or add a particle in these orbitals, and the symbol $\tilde j$ indicates
the time-reversal operation on $j$. 

This wave function is clearly a superposition of components
with different number of particles. Within the BCS theory, one determines the orbitals $\phi_j$, 
the parameters $u_j$ and $v_j$ that appear in (\ref{eq:BCS_wf}), and other relevant quantities, 
by minimisation of the total energy (see Ch. 6 of Ref. \cite{RS.80}). The most relevant quantity is
the pairing gap $\Delta$, that corresponds to the minimal excitation energy of the system and
can be interpreted as the binding energy of the Cooper pairs. The results of BCS can be 
compared with the experimental findings, and the success of the comparison is, in the present context, 
a further example of a symmetry-breaking wave function that describes nuclei more efficiently
than a symmetry-conserving one. The symmetry restoration can be realised in a similar way
as in Eq. (\ref{eq:ang_proj}), through a rotation in an abstract space by the so-called 
gauge angle. The reader is advised to consult Refs. \cite{RS.80,BB.05}. The generalisation of
BCS, namely the Hartree-Fock-Bogoliubov (HFB) theory, is also described in \cite{RS.80}.
Nowadays, the evolution of pairing when going towards the drip lines, or in other dilute systems
like the crust of neutron stars, is under discussion. Another subject which is of great 
interest, but goes beyond the scope of this lecture, is whether pairing between protons and
neutron is strong enough to give rise to a condensate \cite{SBC.16}. 

In summary, space deformation and pairing are both characterised by non-vanishing
expectation values of operators that correspond to the broken symmetries in the intrinsic
system. In the case of quadrupole deformation, the expectation value of the quadrupole moment
$\hat Q$ shows up. One can of course generalise to other deformations like octupole.
In the case of pairing, where the wave function (\ref{eq:BCS_wf}) mixes states 
that differ by two particles, the non-vanishing expectation value is that of 
$\sum_j a_j a_{\tilde j}$ (or its complex conjugate). 

From the DFT viewpoint, the breaking of the rotational symmetry 
implies that the density $\rho(\vec r)$ can assume an intrinsic non-spherical shape, and the
breaking of number symmetry implies that EDFs can also depend on a generalised density
that is called abnormal density (or pairing tensor) and reads $\kappa(\vec r)=
\langle a(\vec r)a(\vec r) \rangle$ (where $a(\vec r)$ is the annihilation operator of a particle at
point $\vec r$ and $a^\dagger(\vec r)$ is the corresponding creation operator). We 
remind the reader that the usual density that we have been dealing with so far reads
$\rho(\vec r)=\langle a^\dagger(\vec r)a(\vec r) \rangle$ in second quantisation. 

A full-fledged formulation of DFT with symmetry breaking goes beyond our scope here. The present
Section can be considered an elementary introduction to specialised lectures on this topic, like
Ref. \cite{LNP.879}.

\section{Extension to the time-dependent case}\label{sec:TD}

The extension of the HK theorem to the time-dependent case has been proposed by 
Runge and Gross in Ref. \cite{RG.84}. The Runge-Gross theorem guarantees that an exact
functional exists, in principle, also for the time-dependent case. More precisely, 
given a system whose behaviour is governed by an Hamiltonian
\begin{equation}
H' = H + v_{\rm pert}(t),
\end{equation} 
where the second term on the r.h.s. is a perturbing time-dependent potential (whose dependence 
on all degrees of freedom of the system is implicit), 
the theorem establishes a one-to-one correspondence between $v_{\rm pert}(t)$ and the 
time-dependent density $\rho(t)$, that is, 
\begin{equation}\label{eq:mapping}
v_{\rm pert}({\vec r}, t) \ \ \ \Leftrightarrow \ \ \ \rho({\vec r}, t). 
\end{equation}
The proof of the Runge-Gross theorem is more subtle and involved than the proof 
of the static HK theorem that we have discussed in Sec. \ref{sec:basics}. 
This theorem constitutes the basis of time-dependent DFT (TDDFT). 
For a pedagogical introduction to the subject, the reader can consult Chapter 4
of Ref. \cite{FNM.03}.

TDDFT can also be approached within the Kohn-Sham scheme. In fact, the time-dependent 
density can be expressed in terms of time-dependent Kohn-Sham orbitals, 
\begin{equation}
\rho({\vec r}, t) = \sum_j \vert \phi_j({\vec r}, t) \vert^2,
\end{equation}
and these orbitals must satisfy time-dependent Schr\"odinger equations of the type
\begin{equation}\label{eq:td1}
i\hbar \frac{\partial}{\partial t}\phi_j(\vec r,t) = \left[ -
\frac{\hbar^2}{2m}\nabla^2 + v_{\rm KS}(\vec r,t) + v_{\rm pert}({\vec r}, t) \right] \phi_j({\vec r},t).
\end{equation}
$v_{\rm KS}(\vec r,t)$ is the sum of three terms. The external potential $v_{\rm ext}({\vec r}, t)$ and the Hartree potential 
$v_{\rm Hartree}({\vec r}, t)$ are the straightforward generalisation
of the first two terms at the l.h.s. of Eq. (\ref{eq:KS}),
where now the density is taken to be time-dependent. It is much less obvious how to
extend the exchange-correlation potential defined in Eq. (\ref{eq:KS}) to the time-dependent case,
that is, how to determine what must be inserted in Eq. (\ref{eq:td1}). 
Causality arguments play a role when one wishes to design a sensible strategy to build
this potential: as pointed out in Ref. \cite{vL.98}, the potential must not feel at time $t$ the changes
that the densities may undergo at later times, that is, at $t' > t$. In fact, the rigorous 
proof of the Runge-Gross theorem implies a further key point that one should grasp, namely that
the mapping (\ref{eq:mapping}) holds {\em for a given 
initial condition at time $t_0$}. In other terms, an exact exchange-correlation functional 
evaluated at time $t$ should have memory of the whole previous history of the system for $t' < t$. 

In electronic systems, the construction of exchange-correlation functionals with memory
is still in its infancy. Most calculations adopt the simple adiabatic approximation, 
in which also the exchange-correlation functional has the same form as the static one, Eq. (\ref{eq:KSfunctional}), 
but is evaluated making use of the time-dependent density, {\em viz.}
\begin{equation}
v^{\rm (ALDA)}_{\rm xc}({\vec r}, t) = v_{\rm xc}[\rho({\vec r},t)], 
\end{equation}
where the label ALDA stays for adiabatic LDA. ALDA is basically the only option that has been
so far explored in the nuclear case. Such approximation
is expected to work better if the time scale of the perturbation is slow.

Given some assumption for $v_{KS}$, one can solve directly the time-dependent Kohn-Sham equations.
A practical way is the following. If $h(t)$ is the sum of $v_{\rm KS}$ plus the kinetic energy, the
time evolution of the density can be written as
\begin{equation}\label{eq:TDHF}
i\hbar \frac{d}{dt}\rho(t) = \left[ h(t), \rho(t) \right].
\end{equation}
Given an initial condition in which $\rho(t=t_0)$ is different from the stationary ground-state
density, this latter equation can be solved by defining a time-evolution operator
$\hat U \equiv e^{-i\frac{\Delta t}{\hbar}\cdot h}$, and by applying it to the density: starting from the density
at time $t_0$, one can in fact write $\rho(t_0+\Delta t)= \hat U \rho(t=t_0)$, where
$\Delta t$ is an appropriate time step. This version of time-dependent DFT has been
widely used. A useful introduction can be found in Refs. \cite{RS.04,Nakatsukasa:2016}, but 
many applications to nuclear vibrations \cite{MRSSS.05}, fusion processes \cite{SKUO.13}, and 
heavy-ion reactions \cite{SY.16}, among others, can be found in the recent literature.

The time-dependent equation (\ref{eq:TDHF}) can be considered in the case of small
external perturbations, namely it can be linearized. This linearization should be appropriate
in the case of nuclear vibrations like the so-called giant resonances, and it is called 
linear response theory in general, and Random Phase Approximation (RPA) in the nuclear
physics context. In this specific case, the adiabatic approximation
should make sense as already pointed out in \cite{RG.84}; whether this approximation describes well 
specific physics cases, is a different and open question. 

The derivation of RPA from
Eq. (\ref{eq:TDHF}) can be found in standard textbooks (see, e.g., Chapter 8 of 
\cite{RS.80}). We will not repeat it in detail here, but rather discuss the key ideas. 
If the perturbation $v_{\rm pert}$ is small, it makes sense to assume it is harmonic,
\begin{equation}\label{eq:vpert}
v_{\rm pert}({\vec r}, t) = v_{\rm pert}({\vec r})e^{i\omega t} + {\rm h.c.}, 
\end{equation}
and then the density will display changes $\delta\rho(\vec r,t)$ that behave similarly as a function of time, that is, 
\begin{eqnarray}\label{eq:rho}
\rho(\vec r,t) & = & \rho(\vec r) + \delta\rho(\vec r,t), \nonumber \\
\delta\rho(\vec r,t) & = & \delta\rho(\vec r) e^{i\omega t} + {\rm h.c.},
\end{eqnarray}
where $\rho(\vec r)$ is here the stationary ground-state density. RPA equations can be obtained by
inserting (\ref{eq:vpert}) and (\ref{eq:rho}) in (\ref{eq:TDHF}), and retaining only terms that are
linear in $\delta\rho$ or in $\delta h \equiv \frac{\delta h}{\delta \rho}\delta\rho$. 
If we go to limit of vanishing external perturbation, and if the final result is expressed on the basis of the Kohn-Sham 
orbitals, it reads
\begin{equation}\label{eq:RPA}
\left( \begin{array}{cc} A & B \\ -B & -A \end{array} \right) \left( \begin{array}{c} X \\ Y 
\end{array} \right) = \hbar\omega \left( \begin{array}{c} X \\ Y 
\end{array} \right).
\end{equation}
This is an eigenvalue equation, and $\hbar\omega$ are the energies of the small-amplitude vibrations.
For each of these modes, we have defined an eigenvector which is made up with the so-called
forward-going and backward-going amplitudes $X_{mi}$ and $Y_{mi}$, respectively\footnote{Here, 
$i,j$ are used to label occupied orbitals and $m,n$ are used to label unoccupied orbitals.}.
$X_{mi}$ ($Y_{mi}$) represents the probability amplitude that a transition from the orbital 
$i$ to $m$ ($m$ to $i$\footnote{Note that introducing the backward-going amplitudes is consistent with
the linearity of the response, but implies relaxing the assumption that the ground-state is made up 
with the auxiliary orbitals filled up to the Fermi energy, and empty above this energy.}) 
contributes to the vibrational mode. The matrix elements that appear in Eq. 
(\ref{eq:RPA}) are defined by
\begin{eqnarray}
A_{mi,nj} & = & \delta_{ij}\delta_{mn}
(\varepsilon_m-\varepsilon_i) + 
\int d^3r\ d^3r^\prime\ \phi^*_m(\vec r)\phi^*_j(\vec r^\prime) \frac{\delta h({\vec r})}
{\delta\rho({\vec r^\prime})} \phi_i(\vec r)\phi_n(\vec r^\prime), \nonumber \\
B_{mi,nj} & = & \int d^3r\ d^3r^\prime\ \phi^*_m(\vec r)\phi^*_n(\vec r^\prime) \frac{\delta h({\vec r})}
{\delta\rho({\vec r^\prime})} \phi_i(\vec r)\phi_j(\vec r^\prime),
\end{eqnarray}
where $\varepsilon$ denotes the energies of the Kohn-Sham orbitals that have been introduced in
Eq. (\ref{eq:KS}). The physical interpretation of $\frac{\delta h({\vec r})}{\delta\rho({\vec r^\prime})}$ is pretty 
intuitive: the changes of the density affect the single-particle 
effective operator $h$ (kinetic energy plus $v_{\rm KS}$) and this, in turn, produces a rearrangement 
of the system. In other words,
the system will perform harmonic oscillations around the ground-state that are governed by the residual force
\begin{equation}\label{eq:vres}
V_{\rm res}({\vec r},{\vec r}') \equiv \frac{\delta h({\vec r})}{\delta\rho({\vec r^\prime})}.
\end{equation}

RPA equations can be also written in real-space instead than on a basis. This
implementation allows a proper treatment of the continuum, but may become quite demanding if one deals
with EDFs with many different terms. In fact, we have only mentioned the total density $\rho$ so far, 
but, according to the discussion of Sec. \ref{sec:examples}, realistic EDFs will bring in dependence
on $\tau$, $\vec\nabla \rho$, $\vec J$ etc. Then, the variation of these densities
will add new terms to the residual interaction (\ref{eq:vres}), like 
$\frac{\delta h({\vec r})}{\delta\tau({\vec r^\prime})}$ and so on. 

A multitude of variants and applications of RPA exist in the literature. Part of it
is reviewed, e.g., in Ref. \cite{BHF.03}. We should add, in this context, that recently 
a new way of implementing RPA has been proposed \cite{Nakatsukasa:07}, which
is not yet discussed in standard textbooks and review papers. This method, called Finite Amplitude Method (FAM) 
realizes the above ideas of linear response in a quite straighforward way and has the advantage that
the numerical implementation is far easier than those of standard methods. 

\section{Examples of RPA calculations}\label{sec:examples_RPA}

The best example of nuclear collective motion are the giant resonances, as we have already briefly
mentioned. They are coherent modes, that dominate the response of nuclei in the energy region around 
$\approx$ 10-30 MeV. Their properties are consistent with the assumption of an elastic
behaviour \cite{Bertsch.74} and, in this respect, their integral properties are well described
by RPA. Giant resonances carry different spatial angular momentum $L$,
spin $S$, and isospin $T$. The states in which nucleons with opposite spin
vibrate in phase (out of phase) are called electric (magnetic) resonances;
if nucleons with opposite isospin, i.e. neutrons and protons,
vibrate in phase (out of phase) one speaks of isoscalar (isovector) resonances.

We will mention only one of the multipoles in what follows. An external electromagnetic field excites the isovector dipole
in a practically exclusive way. The response is dominated by the so-called isovector
giant dipole resonance (IVGDR) at an energy around 80A$^{-1/3}$: at that energy,
the photon wavelength is much larger than the nuclear size and the associated
electric field, that acts on the protons, is constant in space and
thus consistent with a potential that is linear in the proton coordinates $\vec r_p$. 
It is very easy to check that if we take a potential of the type $e\sum_p \vec r_p$, 
and we transform it to the center-of-mass system, it becomes
\begin{equation}\label{eq:op_dipole}
{\hat O}_{\rm IV\ dipole} = e\frac{N}{A} \sum_p \vec r_p -
e\frac{Z}{A} \sum_n \vec r_n.
\end{equation}
In the center-of-mass system, protons and neutrons are displaced in opposite directions and the
strong neutron-proton interaction acts as a restoring force. 

\begin{figure}[t]
\includegraphics[clip=true,width=0.45\linewidth]{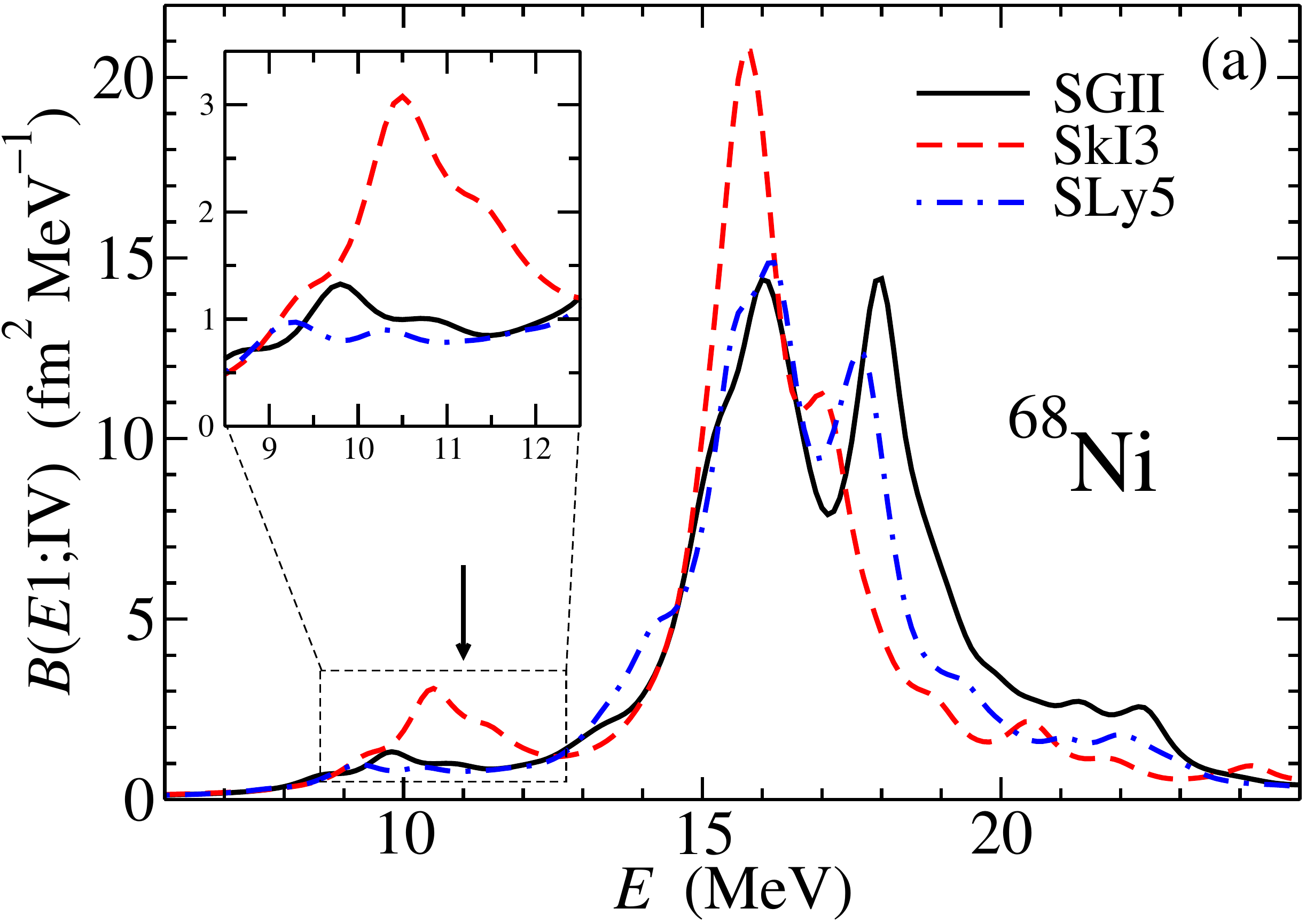}
\includegraphics[clip=true,width=0.45\linewidth]{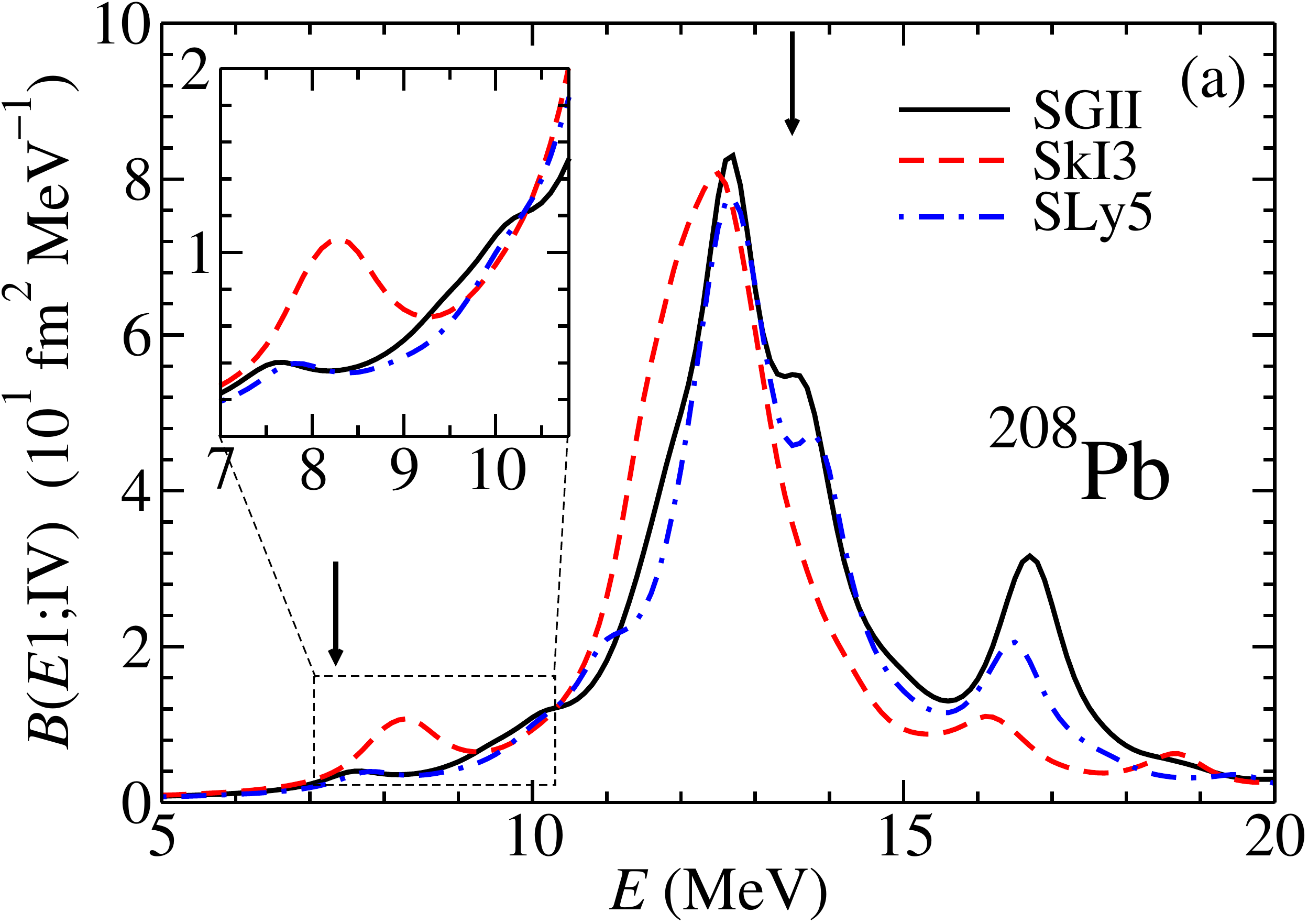}
\\
\includegraphics[clip=true,width=0.45\linewidth]{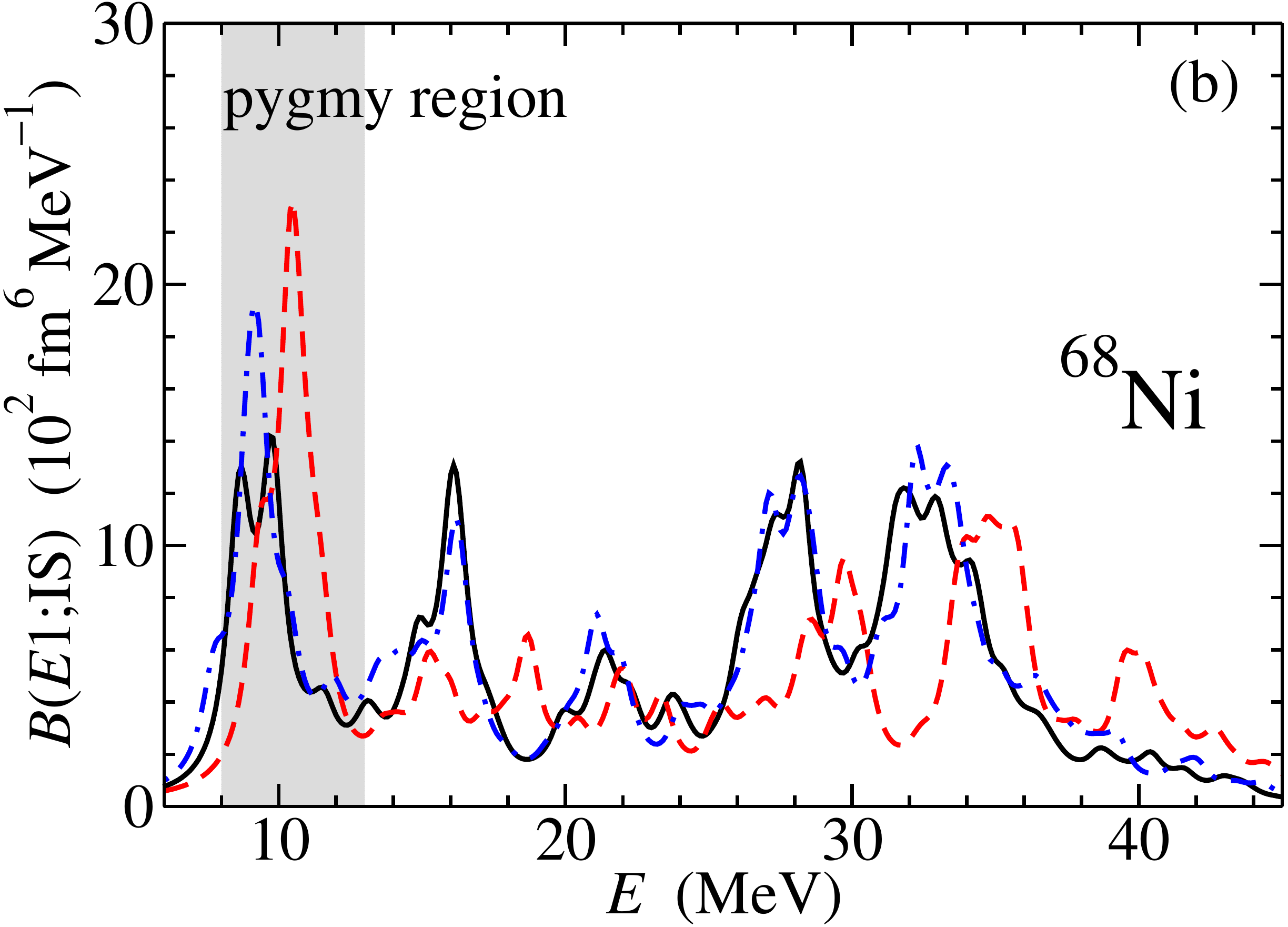}
\includegraphics[clip=true,width=0.45\linewidth]{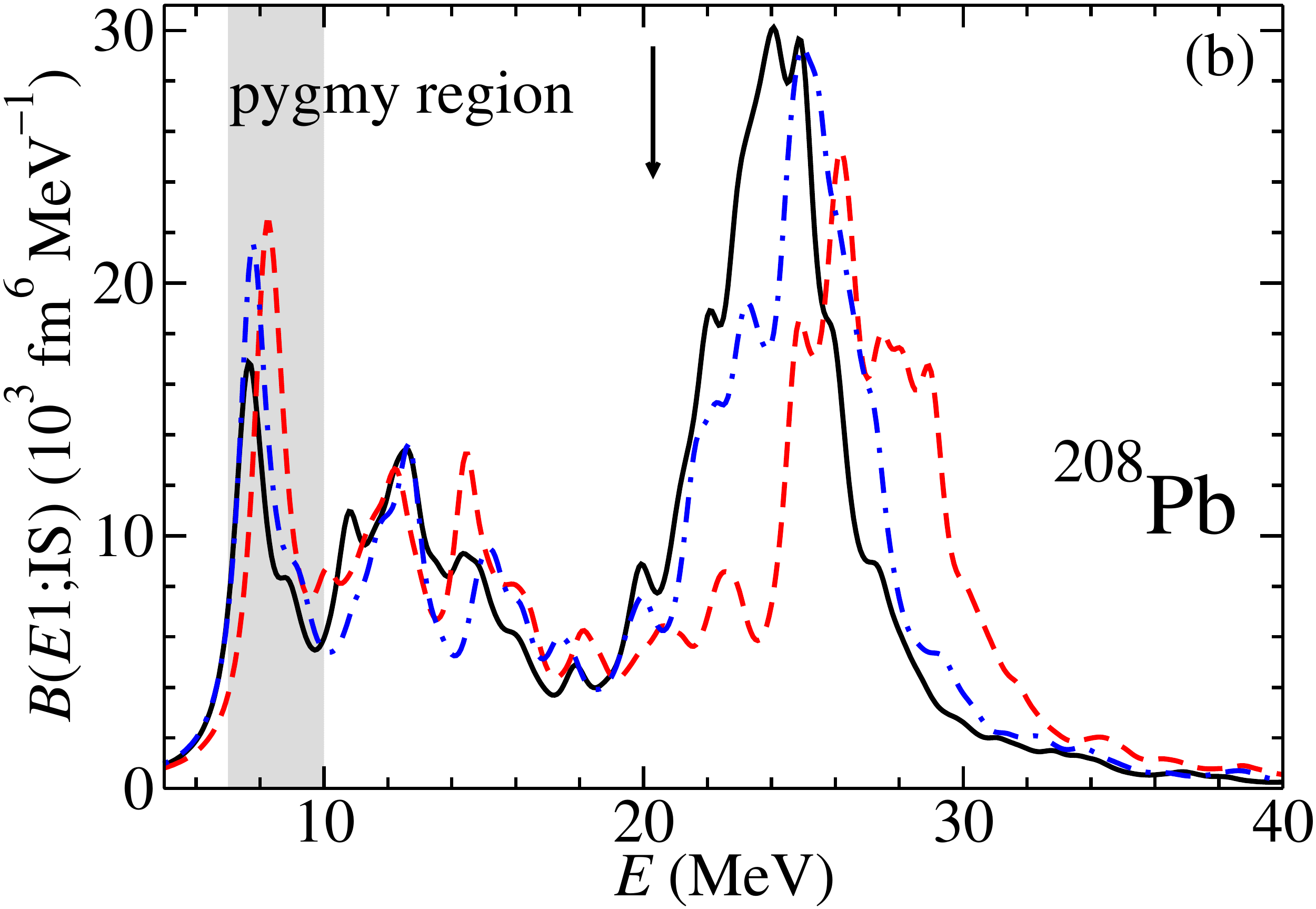}
\caption{Strength functions (\ref{eq:strength}) corresponding to the isovector (a) and isoscalar (b)
dipole response of ${}^{68}$Ni (left panels) and ${}^{208}$Pb (right panels) as a function of the excitation energy. The insets display on a larger scale the low-energy, or pygmy, region. The predictions obtained 
with different Skyrme functionals are depicted (cf. the main text). 
The results from Eq. (\ref{eq:strength}) are averaged with Lorentzian functions having 1 MeV width.
Black arrows indicate the experimental results from Refs. \cite{berm1975,weil2009,ryez2002,garg1999}.
Figure taken from Ref. \cite{rocamaza2012}.}
\label{fig_dipole}
\end{figure}

In Fig. \ref{fig_dipole} we show strength functions resulting from RPA calculations
with different Skyrme functionals (SGII from Ref. \cite{sgii}, SkI3 from Ref. 
\cite{ski3} and SLy5 from Ref. \cite{Cha.98}). Strength functions are defined by
\begin{equation}\label{eq:strength}
S(E) = \sum_n \vert \langle n \vert \hat O \vert 0 \rangle \vert^2 \delta(E-
\hbar\omega_n),
\end{equation}
where $n$ labels the solutions of Eq. (\ref{eq:RPA}), namely $\hbar\omega_n$ is an
eigenvalue and $\vert n\rangle$ is the corresponding eigenvector, $\vert 0 \rangle$ 
is the ground-state, and $\hat O$ is a generic operator. Panels (a) correspond to the
choice of the isovector operator (\ref{eq:op_dipole}), and the main IVGDR peak is compared
with the photoabsorbtion result of Ref. \cite{berm1975}, while the low-lying peak which is
sometimes referred to as ``pygmy'' dipole resonance (PDR) is compared with the experimental
findings of Refs. \cite{weil2009,ryez2002}. Panels (b) show instead the response to the
isoscalar dipole operator,
\begin{equation}
{\hat O}_{\rm IS\ dipole} = \sum_i \left( r_i^3 - \frac{3}{5}\langle r^2\rangle r_i \right) Y_{1M}(\hat r_i).
\end{equation}
Details for this choice can be found in Ref. \cite{rocamaza2012}. The result for the main
resonance excited by this operator can be compared with the result of experiments like inelastic 
($\alpha$,$\alpha'$) scattering, as we have done by taking the finding of \cite{garg1999}. The results
in the figure seem to indicate that the low-lying dipole response has more isoscalar than 
isovector character.

Several textbooks have been devoted to the whole subject of nuclear giant resonances and reader
can consult Refs. \cite{Harakeh_book,BBB_book}. The evolution of collective modes in neutron-rich 
nuclei is trated in Ref. \cite{PVKC.07}.

\section{Limitations of EDFs}\label{sec:limitations}

There is still debate on which are the intrinsic limitations of DFT in nuclear physics, and
which are the practical limitations (due to the specific choices for the form of the EDF, 
and for the protocol to fit the parameters). Disentangling the two aspects is not easy. 

The typical example of such debate is the issue with single-particle states.
As we mentioned at the end of Sec. \ref{sec:MF}, the level density in nuclei is consistent
with a value of the effective mass $m^*/m$ that lies around 0.7, but becomes close to 1 in the 
vicinity of the Fermi surface. Transfer experiments, although subject to a large amount of
ambiguities, can provide access to the single-particle strength distributions in nuclei. Qualitatively,
these experiments have shown that, close to the Fermi surface, these distributions display
dominant peaks, that is, are consistent with the existence of particle-like or hole-like levels; however,
far from the Fermi surface, the single-particle strength is very fragmented. These patterns have
been known for quite some time, and explained with the coupling between HF single-particle states
and nuclear vibrations (particle-vibration coupling or PVC), as discussed at length in 
Ref. \cite{MBBD.85}. However, there are claims that, although PVC may be needed to explain single-particle
fragmentation, the centroids of single-particle states must be obtained through improvements of
current EDFs \cite{Tarpanov:2014}.

Even for observables that are the typical focus of nuclear DFT, like masses, we have mentioned in
Sec. \ref{sec:examples_gs} that existing EDFs have errors that are typically around 
1-2 MeV. Mass models that are more successful, and reduce this error below 1 MeV, include 
some terms that go outside the actual DFT philosophy. These terms correspond to 
vibrational or rotational correlations (cf. \cite{Goriely} and references therein).

There are other open questions that are related to the limits of current DFT implementations.
MR-EDF has been shown to be a theory that is mathematically sound, and it has been successful 
in describing many low-lying nuclear spectra; still, pathologies in the results have shown
up, as discussed in \cite{Duguet} and references therein. These pathologies have been 
associated with density-dependence in the Skyrme EDF, which on the other hand seems unavoidable
if one wants to reconcile bulk properties and the effective mass. This is still a serious 
concern for practitioners. Another open question exists, regarding the pairing sector
of the EDFs. In which way an EDF must depend on the abnormal density $\kappa(\vec r)$, and
possibly on its derivatives etc., is not clear yet. Links with an underlying theory of
pairing in nuclei \cite{Hebeler2009,Idini2012} are still not enough. Proton-neutron pairing, as mentioned at the end of 
Sec. \ref{sec:sym}, is still not fully understood.

Last but not least, as briefly alluded to in Sec. \ref{sec:examples_RPA}, time-dependent DFT in
the RPA implementation is known to be unable to reproduce the width of giant resonances, that
is typically from few to several MeV. RPA can reproduce the so-called escape width, associated 
with particles decaying to the continuum, but not the remaining width, so-called spreading 
width. This letter is known to be accounted for in calculations that include 
PVC \cite{Colo:1994,Yifei,Xavier:2017}. Whether functionals with memory effect may or not improve 
on this, is a further open question.

Several of these examples hint that vibrational correlations, pairing correlations, or other 
kinds of dynamical correlations, although in principle should be included in ``exact'' DFT, 
in practice are not included in existing EDFs. These limitations also call for a more 
serious confrontation between DFT and underlying theories, on the one hand, but also between 
DFT and many-body theories: this is not systematically carried on as it should, in the nuclear
structure domain.

\section{Conclusions}\label{sec:conclu}

We have tried to give the reader a brief survey about DFT, namely about its foundations and about the 
main specific features of nuclear DFT. Due to the scope of this volume, we have tried to give 
priority to pedagogical arguments, while we have simply provided references for all those 
detailed topics that are already well covered in the literature. As it has been stressed 
at the beginning, this contribution is meant to be an advanced, up-to-date introduction
that serves mainly as an orientation for those interested in deepening their understanding of
the large variety of DFT-related approaches.

The hope is to have been able to convince the reader that DFT has a number of advantages. Its basic principle 
is very transparent and intuitive: one writes the total energy as a function of the density, 
and chooses a way to minimise the energy! The HK theorem guarantees, then, that all bulk 
observables can be obtained: the density determines the shape of the system, the expectation 
values of relevant one-body operators (like electric quadrupole, or magnetic dipole, or higher
multipole operators) as well as the total energy. The main advantage is that the applicability of DFT is very broad. 
All nuclei are amenable to a DFT description, except perhaps the very light ones. Moreover, 
there is a natural link with infinite nuclear matter and, thus, with the physics of compact objects 
like neutron stars.

After discussing DFT in electronic systems, we have, in fact, started our discussion by introducing nuclear DFT 
in uniform, symmetric nuclear matter. A sensible theory of such prototype system must display 
a mechanism for saturation; a good reproduction of the saturation point ($\rho_0$ = 0.16 
fm$^{-3}$, $e_0$ = $-16$ MeV) is a necessary condition to describe also finite nuclei in a 
reasonable way. We have, then, provided arguments why na\"ive mean-field cannot be, in this
respect, a sensible theory while DFT can. Of course, realistic EDFs should not simply
provide a mechanism of nuclear saturation but should be flexible enough to account for
surface properties, spin and isospin properties, and so on. Accordingly, we have discussed the 
appearance of different kinds of densities in existing EDFs.

A specific feature of finite nuclei is the appearance of spontaneous symmetry-breaking and
restoration. Although specialised lectures exist on those topics, we have tried to
give here an introduction through simple examples and arguments. In the case of
deformed nuclei, the rotational symmetry is broken in the intrinsic reference frame, and
rotations in the laboratory frame restore the symmetry. Within DFT, that has necessarily to
be formulated in the intrinsic frame, as we have argued in Sec. \ref{sec:intrinsic}, 
this implies the appearance of densities that are not
rotationally invariant (that is, quadrupole or octupole moments as we have just stressed).
A more intriguing example is that of pairing correlations that imply, instead, the appearance
of the abnormal density or pairing tensor in the EDFs, breaking the particle number symmetry.

Last but not least, we have discussed the time-dependent case. The time-dependent DFT
is still amenable to many improvements. Essentially all groups are using, so far, the adiabatic
approximation. While this approximation, that corresponds to RPA in the nuclear
structure theory language, has been successful in reproducing the main properties
of the nuclear collective motion like the giant resonances, certainly examples
of large-amplitude, slow motions exist in the nuclear case and call for more
advanced approaches. 

There are topics that we have not dealt with in this lecture and/or topics that have not been
tackled yet by practitioners. One main issue is that, while in several papers the quest
for a universal energy functional is advocated, still many different implementations
of EDFs and even more variants based on different parameter sets are available on the 
market. The situation in the case of electronic systems is, anyway, not much better \cite{Burke.12}.
Not only this fact makes DFT a domain that is believed to be for experts only but, more
importantly, it raises the question if EDFs are systematically improvable or not. 
EFT has been proposed as a way out to this dilemma. However, a full derivation of DFT
from EFT is not available yet.

Another aspect is the confrontation between DFT and many-body theory. The mapping of
some many-body approximation into an EDF shoud be pursued more systematically, and is expected
also to shed light on the present limitations of nuclear DFT. We have briefly mentioned the
fact that DFT cannot describe the fragmentation of single-particle or collective states.
This has been traditionally explained, in the nuclear context, by the coupling of
single-particle and collective degrees of freedom (vibrations or rotations). Merging
this description with up-to-date DFT, still represents a challenging task. 

In summary, we may like to quote a saying by W. Kohn, and state that ``DFT is an 
exactification of HF''; however, even though this statement may be true in principle, in practice many
steps are still required to make the DFT description of the rich nuclear phenomenology 
more and more accurate. 

\acknowledgments
This work is also the result of many collaborations during the last decades. The author would like to
thank, for the common work and the beneficial discussions on the topics that are dealt with 
in this paper, K. Bennaceur, P.F. Bortignon, J. Dobaczewski, T. Duguet, M. Grasso, E. Khan, D. Lacroix, H. Liang, J. Margueron, 
J. Meyer, Y. Niu, 
P.-G. Reinhard, 
P. Ring, X. Roca-Maza, H. Sagawa, N. Van Giai, E. Vigezzi, and D. Vretenar.
The author also expresses his thanks to G. Accorto, M. Barani, R. Romano for their reading of
the manuscript and help in the preparation of figures. Funding from the European Union's Horizon 2020 research and innovation
programme under grant agreement No 654002 is also acknowledged.

\bibliography{bibliography}

\begin{thebibliography}{100}
\expandafter\ifx\csname url\endcsname\relax\def\url#1{\texttt{#1}}\fi
\expandafter\ifx\csname urlprefix\endcsname\relax\def\urlprefix{URL }\fi

\bibitem{Thoennessen:2017}
\NAME{Thoennessen M.}, \IN{International Journal of Modern Physics
  E}{26}{2017}{1730003}.

\bibitem{CH.08}
\NAME{Chamel N. \atque Haensel P.}, \IN{Living Reviews in
  Relativity}{11}{2008}{10}.

\bibitem{Inoue:2013}
\NAME{Inoue T., Aoki S., Doi T., Hatsuda T., Ikeda Y., Ishii N., Murano K.,
  Nemura H. \atque Sasaki K.}, \IN{Phys. Rev. Lett.}{111}{2013}{112503}.

\bibitem{Weinberg:1979}
\NAME{Weinberg S.}, \IN{Physica A: Statistical Mechanics and its
  Applications}{96}{1979}{327}.

\bibitem{RevModPhys.81.1773}
\NAME{Epelbaum E., Hammer H.-W. \atque Mei\ss{}ner U.-G.}, \IN{Rev. Mod.
  Phys.}{81}{2009}{1773}.

\bibitem{Gandolfi2016}
\NAME{Lynn J.~E., Tews I., Carlson J., Gandolfi S., Gezerlis A., Schmidt K.~E.
  \atque Schwenk A.}, \IN{Phys. Rev. Lett.}{116}{2016}{062501}.

\bibitem{Simonis2017}
\NAME{Simonis J., Stroberg S.~R., Hebeler K., Holt J.~D. \atque Schwenk A.},
  \IN{Phys. Rev. C}{96}{2017}{014303}.

\bibitem{RS.80}
\NAME{Ring P. \atque Schuck P.}, \TITLE{The Nuclear Many-Body Problem}
  (Springer) 1980.

\bibitem{Bertulani}
\NAME{Bertulani C.}, \TITLE{Nuclear Physics in a Nutshell} (Princeton
  University Press) 2007.

\bibitem{ParrYang}
\NAME{Parr R.~G. \atque Yang W.}, \TITLE{Density Functional Theory of Atoms and
  Molecules} (Oxford University Press) 1989.

\bibitem{DG.90}
\NAME{Dreizler R. \atque Gross E.}, \TITLE{Density Functional Theory}
  (Springer) 1990.

\bibitem{Gross:1996}
\NAME{Gross E., Dobson J. \atque Petersilka M.}, \TITLE{Density Functional
  Theory} (Springer) 1996.

\bibitem{Burke.12}
\NAME{Burke K.}, \IN{The Journal of Chemical Physics}{136}{2012}{150901}.

\bibitem{Becke.14}
\NAME{Becke A.~D.}, \IN{The Journal of Chemical Physics}{140}{2014}{18A301}.

\bibitem{BHF.03}
\NAME{Bender M., Heenen P.-H. \atque Reinhard P.-G.}, \IN{Rev. Mod.
  Phys.}{75}{2003}{121}.

\bibitem{LNP.641}
\NAME{Lalazissis G., Ring P. \atque Vretenar D.} (Eds.), \TITLE{Extended
  Density Functionals in Nuclear Structure Physics. Lecture Notes in Physics.
  Vol. 641} (Springer) 2004.

\bibitem{LNP.879}
\NAME{Duguet T.}, \TITLE{Lecture Notes in Physics. Vol. 879} (Springer) 2004,
  p. 293.

\bibitem{Doba:archive}
\NAME{Dobaczewski J.}, \TITLE{Interactions, symmetry breaking, and effective
  fields from quarks to nuclei. (a primer in nuclear theory)},
  arXiv:nucl-th/0301069 (2003).

\bibitem{VALR.05}
\NAME{Vretenar D., Afanasjev A., Lalazissis G. \atque Ring P.}, \IN{Physics
  Reports}{409}{2005}{101}.

\bibitem{NVR.11}
\NAME{Nik\ifmmode \check{s}\else \v{s}\fi{}i\ifmmode~\acute{c}\else \'{c}\fi{}
  T., Vretenar D. \atque Ring P.}, \IN{Progress in Particle and Nuclear
  Physics}{66}{2011}{519}.

\bibitem{HK.64}
\NAME{Hohenberg P. \atque Kohn W.}, \IN{Phys. Rev.}{136}{1964}{B864}.

\bibitem{KS.65}
\NAME{Kohn W. \atque Sham L.~J.}, \IN{Phys. Rev.}{140}{1965}{A1133}.

\bibitem{BCOR.04}
\NAME{Broglia R.~A., Col\`o G., Onida G. \atque Roman H.~E.}, \TITLE{Solid
  State Physics of Finite Systems} (Springer) 2004.

\bibitem{Fetter}
\NAME{Fetter A.~L. \atque Walecka J.~D.}, \TITLE{Quantum Theory of
  Many-Particle Systems} (McGraw-Hill) 1971.

\bibitem{Mattuck}
\NAME{Mattuck R.~D.}, \TITLE{A Guide to Feynman Diagrams in the Many-Body
  Problem} (Dover) 1976.

\bibitem{PW.81}
\NAME{Perdew J.~P. \atque Zunger A.}, \IN{Phys. Rev. B}{23}{1981}{5048}.

\bibitem{PS.01}
\NAME{Perdew J.~P. \atque Schmidt K.}, \IN{AIP Conference
  Proceedings}{577}{2001}{1}.

\bibitem{BM1}
\NAME{Bohr A. \atque Mottelson B.~R.}, \TITLE{Nuclear Structure}, Vol.~I
  (Benjamin, New York) 1969.

\bibitem{BM2}
\NAME{Bohr A. \atque Mottelson B.~R.}, \TITLE{Nuclear Structure}, Vol.~II
  (Benjamin, New York) 1975.

\bibitem{Migdal}
\NAME{Migdal A.~B.}, \TITLE{Theory of Finite Fermi Systems and Applications to
  Atomic Nuclei} (Wiley) 1967.

\bibitem{Solovev}
\NAME{Soloviev V.~G.}, \TITLE{Theory of Atomic Nuclei: Quasiparticles and
  Phonons} (Institute of Physics Publishing) 1992.

\bibitem{GC.tbp}
\NAME{Garg U. \atque Col\`o G.}, arXiv:nucl-exp/1801.03672, and {P}rogress in
  Particle and Nuclear Physics (in press) (2018).

\bibitem{topical_volume}
\NAME{{Li, Bao-An}, {Ramos,\`Angels}, {Verde, Giuseppe} \atque {Vida\~na,
  Isaac} E.}, \IN{Eur. Phys. J. A}{50}{2014}{}.

\bibitem{Lattimer:2013}
\NAME{Lattimer J.~M. \atque Lim Y.}, \IN{The Astrophysical
  Journal}{771}{2013}{51}.

\bibitem{Li:2013}
\NAME{Li B.-A. \atque Han X.}, \IN{Physics Letters B}{727}{2013}{276}.

\bibitem{oertel2017}
\NAME{Oertel M., Hempel M., Kl\"ahn T. \atque Typel S.}, \IN{Rev. Mod.
  Phys.}{89}{2017}{015007}.

\bibitem{Horowitz:2014}
\NAME{Horowitz C.~J., Brown E.~F., Kim Y., Lynch W.~G., Michaels R., Ono A.,
  Piekarewicz J., Tsang M.~B. \atque Wolter H.~H.}, \IN{Journal of Physics G:
  Nuclear and Particle Physics}{41}{2014}{093001}.

\bibitem{Gogny:1975}
\NAME{Gogny D.}, \IN{Nuclear Physics A}{237}{1975}{399}.

\bibitem{SerotWalecka}
\NAME{Serot B. \atque Walecka J.}, \IN{Advances in Nuclear
  Physics}{16}{1986}{1}.

\bibitem{Sel.14}
\NAME{Sellahewa R. \atque Rios A.}, \IN{Phys. Rev. C}{90}{2014}{054327}.

\bibitem{Accorto}
\NAME{Accorto G.}, Master's thesis, University of Milano (2015).

\bibitem{BB.67}
\NAME{Brink D. \atque Boeker E.}, \IN{Nuclear Physics A}{91}{1967}{1}.

\bibitem{Blaizot1980}
\NAME{Blaizot J.~P.}, \IN{Phys. Rep.}{64}{1980}{171}.

\bibitem{Rancati}
\NAME{Rancati A.}, Master's thesis, University of Milano (2016).

\bibitem{MBBD.85}
\NAME{Mahaux C., Bortignon P., Broglia R. \atque Dasso C.}, \IN{Physics
  Reports}{120}{1985}{1}.

\bibitem{Weisskopf}
\NAME{Weisskopf V.~F.}, \IN{Nuclear Physics}{3}{1957}{423}.

\bibitem{Nakatsukasa:2016}
\NAME{Nakatsukasa T., Matsuyanagi K., Matsuo M. \atque Yabana K.}, \IN{Rev.
  Mod. Phys.}{88}{2016}{045004}.

\bibitem{1712.03003}
\NAME{Davesne D., Navarro J., Meyer J., Bennaceur K. \atque Pastore A.},
  \IN{Phys. Rev. C}{97}{2018}{044304}.

\bibitem{Romano}
\NAME{Romano R.}, Master's thesis, University of Milano (2015).

\bibitem{De.80}
\NAME{Decharg\'e J. \atque Gogny D.}, \IN{Phys. Rev. C}{21}{1980}{1568}.

\bibitem{BGG.91}
\NAME{Berger J., Girod M. \atque Gogny D.}, \IN{Computer Physics
  Communications}{63}{1991}{365}.

\bibitem{VB.72}
\NAME{Vautherin D. \atque Brink D.~M.}, \IN{Phys. Rev. C}{5}{1972}{626}.

\bibitem{Cha.97}
\NAME{Chabanat E., Bonche P., Haensel P., Meyer J. \atque Schaeffer R.},
  \IN{Nuclear Physics A}{627}{1997}{710}.

\bibitem{Cha.98}
\NAME{Chabanat E., Bonche P., Haensel P., Meyer J. \atque Schaeffer R.},
  \IN{Nuclear Physics A}{635}{1998}{231}.

\bibitem{RF.95}
\NAME{Reinhard P.-G. \atque Flocard H.}, \IN{Nuclear Physics
  A}{584}{1995}{467}.

\bibitem{Kortelainen2014}
\NAME{Kortelainen M., McDonnell J., Nazarewicz W., Olsen E., Reinhard P.-G.,
  Sarich J., Schunck N., Wild S.~M., Davesne D., Erler J. \atque Pastore A.},
  \IN{Phys. Rev. C}{89}{2014}{054314}.

\bibitem{Engel:1975}
\NAME{Engel Y., Brink D., Goeke K., Krieger S. \atque Vautherin D.},
  \IN{Nuclear Physics A}{249}{1975}{215}.

\bibitem{DD.95}
\NAME{Dobaczewski J. \atque Dudek J.}, \IN{Phys. Rev. C}{52}{1995}{1827}.

\bibitem{DD.96}
\NAME{Dobaczewski J. \atque Dudek J.}, \IN{Acta Physica Polonica
  B}{27}{1996}{95}.

\bibitem{DD.er}
\NAME{Dobaczewski J. \atque Dudek J.}, \IN{Phys. Rev. C}{55}{1997}{3177}.

\bibitem{Raimondi2011}
\NAME{Raimondi F., Carlsson B.~G. \atque Dobaczewski J.}, \IN{Phys. Rev.
  C}{83}{2011}{054311}.

\bibitem{Becker1}
\NAME{Becker P., Davesne D., Meyer J., Pastore A. \atque Navarro J.},
  \IN{Journal of Physics G: Nuclear and Particle Physics}{42}{2015}{034001}.

\bibitem{Becker2}
\NAME{Davesne D., Navarro J., Becker P., Jodon R., Meyer J. \atque Pastore A.},
  \IN{Phys. Rev. C}{91}{2015}{064303}.

\bibitem{Nakada2008}
\NAME{Nakada H.}, \IN{Phys. Rev. C}{78}{2008}{054301}.

\bibitem{FTZ.94}
\NAME{Fayans S., Trykov E. \atque Zawischa D.}, \IN{Nuclear Physics
  A}{568}{1994}{523}.

\bibitem{Fayans1998}
\NAME{Fayans S.~A.}, \IN{Journal of Experimental and Theoretical Physics
  Letters}{68}{1998}{169}.

\bibitem{BCPM}
\NAME{Baldo M., Schuck P. \atque Vi{\~{n}}as X.}, \IN{Physics Letters
  B}{663}{2008}{390}.

\bibitem{Lunney}
\NAME{Lunney D., Pearson J.~M. \atque Thibault C.}, \IN{Rev. Mod.
  Phys.}{75}{2003}{1021}.

\bibitem{Goriely}
\NAME{Goriely S., Chamel N. \atque Pearson J.~M.}, \IN{Phys. Rev.
  C}{93}{2016}{034337}.

\bibitem{Kortelainen:2010}
\NAME{Kortelainen M., Lesinski T., Mor\'e J., Nazarewicz W., Sarich J., Schunck
  N., Stoitsov M.~V. \atque Wild S.}, \IN{Phys. Rev. C}{82}{2010}{024313}.

\bibitem{rel_masses}
\NAME{Nik\ifmmode \check{s}\else \v{s}\fi{}i\ifmmode~\acute{c}\else \'{c}\fi{}
  T., Vretenar D. \atque Ring P.}, \IN{Phys. Rev. C}{78}{2008}{034318}.

\bibitem{Erler2012}
\NAME{Erler J., Birge N., Kortelainen M., Nazarewicz W., Olsen E., Perhac A.~M.
  \atque Stoitsov M.}, \IN{Nature}{486}{2012}{509}.

\bibitem{Afanasjev2013}
\NAME{Afanasjev A., Agbemava S., Ray D. \atque Ring P.}, \IN{Physics Letters
  B}{726}{2013}{680}.

\bibitem{Engel:2007}
\NAME{Engel J.}, \IN{Phys. Rev. C}{75}{2007}{014306}.

\bibitem{Barnea:2007}
\NAME{Barnea N.}, \IN{Phys. Rev. C}{76}{2007}{067302}.

\bibitem{MBS:2009}
\NAME{Messud J., Bender M. \atque Suraud E.}, \IN{Phys. Rev.
  C}{80}{2009}{054314}.

\bibitem{Goutte}
\NAME{Goutte H.}, \TITLE{Symmetry breaking and symmetry restoration in
  mean-field based approached}, in \TITLE{Ecole Joliot Curie}, IN2P3-CNRS
  (C.E.N.B.G., http://www.cenbg.in2p3.fr/joliot-curie/) 2010.

\bibitem{Lacroix}
\NAME{Lacroix D.}, \TITLE{Review of mean-field theory}, in \TITLE{Ecole Joliot
  Curie}, IN2P3-CNRS (C.E.N.B.G., http://www.cenbg.in2p3.fr/joliot-curie/)
  2011.

\bibitem{BN.96}
\NAME{Butler P.~A. \atque Nazarewicz W.}, \IN{Rev. Mod. Phys.}{68}{1996}{349}.

\bibitem{Gaffney2013}
\NAME{Gaffney L.~P., Butler P.~A., Scheck M., Hayes A.~B., Wenander F., Albers
  M., Bastin B., Bauer C., Blazhev A., B{\"o}nig S., Bree N., Cederk{\"a}ll J.,
  Chupp T., Cline D., Cocolios T.~E., Davinson T., De~Witte H., Diriken J.,
  Grahn T., Herzan A., Huyse M., Jenkins D.~G., Joss D.~T., Kesteloot N., Konki
  J., Kowalczyk M., Kr{\"o}ll T., Kwan E., Lutter R., Moschner K., Napiorkowski
  P., Pakarinen J., Pfeiffer M., Radeck D., Reiter P., Reynders K., Rigby
  S.~V., Robledo L.~M., Rudigier M., Sambi S., Seidlitz M., Siebeck B., Stora
  T., Thoele P., Van~Duppen P., Vermeulen M.~J., von Schmid M., Voulot D., Warr
  N., Wimmer K., Wrzosek-Lipska K., Wu C.~Y. \atque Zielinska M.},
  \IN{Nature}{497}{2013}{199}.

\bibitem{BB.05}
\NAME{Brink D.~M. \atque Broglia R.~A.}, \TITLE{Nuclear Superfluidity: Pairing
  in Finite Systems} (Cambridge University Press) 2005.

\bibitem{SBC.16}
\NAME{Sagawa H., Bai C.~L. \atque Col\`o G.}, \IN{Physica
  Scripta}{91}{2016}{083011}.

\bibitem{RG.84}
\NAME{Runge E. \atque Gross E. K.~U.}, \IN{Phys. Rev. Lett.}{52}{1984}{997}.

\bibitem{FNM.03}
\NAME{Fiolhais C., Nogueira F. \atque Marques M.} (Eds.), \TITLE{A primer in
  Density Functional Theory. Lecture Notes in Physics. Vol. 620} (Springer)
  2003.

\bibitem{vL.98}
\NAME{van Leeuwen R.}, \IN{Phys. Rev. Lett.}{80}{1998}{1280}.

\bibitem{RS.04}
\NAME{Reinhard P. \atque Suraud E.}, \TITLE{Introduction to Cluster Dynamics}
  (Wiley) 2004.

\bibitem{MRSSS.05}
\NAME{Maruhn J.~A., Reinhard P.~G., Stevenson P.~D., Stone J.~R. \atque Strayer
  M.~R.}, \IN{Phys. Rev. C}{71}{2005}{064328}.

\bibitem{SKUO.13}
\NAME{Simenel C., Keser R., Umar A.~S. \atque Oberacker V.~E.}, \IN{Phys. Rev.
  C}{88}{2013}{024617}.

\bibitem{SY.16}
\NAME{Sekizawa K. \atque Yabana K.}, \IN{Phys. Rev. C}{93}{2016}{054616}.

\bibitem{Nakatsukasa:07}
\NAME{Nakatsukasa T., Inakura T. \atque Yabana K.}, \IN{Phys. Rev.
  C}{76}{2007}{024318}.

\bibitem{Bertsch.74}
\NAME{Bertsch G.}, \IN{Annals of Physics}{86}{1974}{138}.

\bibitem{berm1975}
\NAME{Berman B.~L. \atque Fultz S.~C.}, \IN{Rev. Mod. Phys.}{47}{1975}{713}.

\bibitem{weil2009}
\NAME{Wieland O., Bracco A., Camera F., Benzoni G., Blasi N., Brambilla S.,
  Crespi F. C.~L., Leoni S., Million B., Nicolini R., Maj A., Bednarczyk P.,
  Grebosz J., Kmiecik M., Meczynski W., Styczen J., Aumann T., Banu A., Beck
  T., Becker F., Caceres L., Doornenbal P., Emling H., Gerl J., Geissel H.,
  Gorska M., Kavatsyuk O., Kavatsyuk M., Kojouharov I., Kurz N., Lozeva R.,
  Saito N., Saito T., Schaffner H., Wollersheim H.~J., Jolie J., Reiter P.,
  Warr N., deAngelis G., Gadea A., Napoli D., Lenzi S., Lunardi S., Balabanski
  D., LoBianco G., Petrache C., Saltarelli A., Castoldi M., Zucchiatti A.,
  Walker J. \atque B\"urger A.}, \IN{Phys. Rev. Lett.}{102}{2009}{092502}.

\bibitem{ryez2002}
\NAME{Ryezayeva N., Hartmann T., Kalmykov Y., Lenske H., von Neumann-Cosel P.,
  Ponomarev V.~Y., Richter A., Shevchenko A., Volz S. \atque Wambach J.},
  \IN{Phys. Rev. Lett.}{89}{2002}{272502}.

\bibitem{garg1999}
\NAME{Garg U.}, \IN{Nuclear Physics A}{649}{1999}{66}.

\bibitem{rocamaza2012}
\NAME{Roca-Maza X., Pozzi G., Brenna M., Mizuyama K. \atque Col\`o G.},
  \IN{Phys. Rev. C}{85}{2012}{024601}.

\bibitem{sgii}
\NAME{Giai N.~V. \atque Sagawa H.}, \IN{Physics Letters B}{106}{1981}{379}.

\bibitem{ski3}
\NAME{Reinhard P.-G. \atque Flocard H.}, \IN{Nuclear Physics
  A}{584}{1995}{467}.

\bibitem{Harakeh_book}
\NAME{Harakeh M.~N. \atque van~der Woude A.}, \TITLE{Giant Resonances
  Fundamental High-Frequency Modes of Nuclear Excitation} (Oxford University
  Press, New York) 2001.

\bibitem{BBB_book}
\NAME{Bortignon P.~F., Bracco A. \atque Broglia R.~A.}, \TITLE{Giant
  Resonances: Nuclear Structure at Finite Temperature} (Harwood) 1999.

\bibitem{PVKC.07}
\NAME{Paar N., Vretenar D., Khan E. \atque Col\`o G.}, \IN{Reports on Progress
  in Physics}{70}{2007}{691}.

\bibitem{Tarpanov:2014}
\NAME{Tarpanov D., Dobaczewski J., Toivanen J. \atque Carlsson B.~G.},
  \IN{Phys. Rev. Lett.}{113}{2014}{252501}.

\bibitem{Duguet}
\NAME{Duguet T., Bender M., Bennaceur K., Lacroix D. \atque Lesinski T.},
  \IN{Phys. Rev. C}{79}{2009}{044320}.

\bibitem{Hebeler2009}
\NAME{Hebeler K., Duguet T., Lesinski T. \atque Schwenk A.}, \IN{Phys. Rev.
  C}{80}{2009}{044321}.

\bibitem{Idini2012}
\NAME{Idini A., Barranco F. \atque Vigezzi E.}, \IN{Phys. Rev.
  C}{85}{2012}{014331}.

\bibitem{Colo:1994}
\NAME{Col\`o G., Van~Giai N., Bortignon P.~F. \atque Broglia R.~A.}, \IN{Phys.
  Rev. C}{50}{1994}{1496}.

\bibitem{Yifei}
\NAME{Niu Y.~F., Col\`o G., Vigezzi E., Bai C.~L. \atque Sagawa H.}, \IN{Phys.
  Rev. C}{94}{2016}{064328}.

\bibitem{Xavier:2017}
\NAME{Roca-Maza X., Niu Y.~F., Col\`o G. \atque Bortignon P.~F.}, \IN{Journal
  of Physics G: Nuclear and Particle Physics}{44}{2017}{044001}.

\end{thebibliography}

\end{document}